\documentclass[%
 reprint,
superscriptaddress,
 amsmath,amssymb,
 aps,
pra,
]{revtex4-2}

\usepackage{multirow}
\usepackage[T1]{fontenc}
\usepackage[colorlinks, linkcolor=red, anchorcolor=red, citecolor=red]{hyperref}
\usepackage{amsthm}
\usepackage{graphicx}
\usepackage{dcolumn}
\usepackage{bm}
\usepackage{float}
\usepackage{algpseudocodex}
\usepackage[linesnumbered, boxed, ruled]{algorithm2e}
\usepackage{caption}
\usepackage{subcaption}
\usepackage{xspace}
\usepackage{quantikz}
\usepackage{tikz,amsmath, amssymb,bm,color}
\usetikzlibrary{calc}

\newcommand{\gTj}{\gate{T_j}}
\newcommand{\gUtheta}{\gate[wires=2, style={fill=white}]{U_{\theta}}}
\newcommand{\gUthetaStar}{\gate[wires=3, style={fill=white}]{U_{\theta}^*}}
\newcommand{\gTiH}{\gate{T_i^H}}
\newcommand{\gJ}{\gate{J}}
\newcommand{\gB}{\gate{B}}
\newcommand{\gC}{\gate{C}}
\newcommand{\gH}{\gate{H}}
\newcommand{\gLn}{\gate{L_n}}
\newcommand{\gKn}{\gate{K_n}}
\newcommand{\gLOne}{\gate{L_1}}
\newcommand{\gKOne}{\gate{K_1}}
\newcommand{\gPn}{\gate[wires=5, style={minimum width=1.3cm, inner xsep=8pt, fill=white}]{P_n}} 

\newcommand{\gFtnp}{\gate[wires=6, style={fill=white,minimum width=1.3cm}]{F^T_{n+1}}}

\newcommand{\gdot}{\vdots{}}
\newcommand{\gRz}{\gate{R_z(\theta)}}
\newcommand{\gHdots}{\qw \; \dots \; \qw}

\newcommand{\gMeasure}{\meter{}} 

\newcommand{\OPQROM}{PolyQROM\xspace}

\graphicspath{{figure/}}

\begin{document}

\preprint{APS/123-QED}

\title{PolyQROM: Orthogonal-Polynomial-Based Quantum Reduced-Order Model for Flow Field Analysis}

\author{Yu Fang}
\affiliation{Institute of Artificial Intelligence, Hefei Comprehensive National Science Center, Hefei, Anhui, 230088, P. R. China}
\affiliation{Institute of Advanced Technology, University of Science and Technology of China, Hefei, Anhui, 230088, P. R. China}
\author{Cheng Xue}
\thanks{xcheng@iai.ustc.edu.cn}
\affiliation{Institute of Artificial Intelligence, Hefei Comprehensive National Science Center, Hefei, Anhui, 230088, P. R. China}
\author{Tai-Ping Sun}
\affiliation{Key Laboratory of Quantum Information, University of Science and Technology of China, Hefei 230026, China}
\author{Xiao-Fan Xu}
\affiliation{Key Laboratory of Quantum Information, University of Science and Technology of China, Hefei 230026, China}
\author{Xi-Ning Zhuang}
\affiliation{Key Laboratory of Quantum Information, University of Science and Technology of China, Hefei 230026, China}
\author{Yun-Jie Wang}
\affiliation{Institute of Advanced Technology, University of Science and Technology of China, Hefei, Anhui, 230088, P. R. China}
\author{Chuang-Chao Ye}
\affiliation{Origin Quantum Computing Company Limited, Hefei, Anhui, 230088, P. R. China}
\author{Teng-Yang Ma}
\affiliation{Origin Quantum Computing Company Limited, Hefei, Anhui, 230088, P. R. China}
\author{Jia-Xuan Zhang}
\affiliation{Key Laboratory of Quantum Information, University of Science and Technology of China, Hefei 230026, China}
\author{Huan-Yu Liu}
\affiliation{Key Laboratory of Quantum Information, University of Science and Technology of China, Hefei 230026, China}
\author{Yu-Chun Wu}
\affiliation{Key Laboratory of Quantum Information, University of Science and Technology of China, Hefei 230026, China}
\affiliation{Institute of Artificial Intelligence, Hefei Comprehensive National Science Center, Hefei, Anhui, 230088, P. R. China}
\author{Zhao-Yun Chen}
\thanks{chenzhaoyun@iai.ustc.edu.cn}
\affiliation{Institute of Artificial Intelligence, Hefei Comprehensive National Science Center, Hefei, Anhui, 230088, P. R. China}
\author{Guo-Ping Guo}
\affiliation{Key Laboratory of Quantum Information, University of Science and Technology of China, Hefei 230026, China}
\affiliation{Origin Quantum Computing Company Limited, Hefei, Anhui, 230088, P. R. China}
\affiliation{Institute of Artificial Intelligence, Hefei Comprehensive National Science Center, Hefei, Anhui, 230088, P. R. China}

\begin{abstract}

Quantum computing promises exponential acceleration for fluid flow simulations, yet the measurement overhead required to extract flow features from quantum-encoded flow field data fundamentally undermines this advantage--a critical challenge termed the ``output problem''. To address this, we propose an orthogonal-polynomial-based quantum reduced-order model (\OPQROM) that integrates orthogonal polynomial basis transformations with variational quantum circuits (VQCs). \OPQROM employs optimized polynomial-based quantum operations to compress flow field data into low-dimensional representations while preserving essential features, enabling efficient quantum or classical post-processing for tasks like reconstruction and classification. By leveraging the mathematical properties of orthogonal polynomials, the framework enhances circuit expressivity and stabilizes training compared to conventional hardware-efficient VQCs. Numerical experiments demonstrate PolyQROM’s effectiveness in reconstructing flow fields with high fidelity and classifying flow patterns with accuracy surpassing classical methods and quantum benchmarks, all while reducing computational complexity and parameter counts. The work bridges quantum simulation outputs with practical fluid analysis, addressing the ``output problem'' through efficient reduced-order modeling tailored for quantum-encoded flow data, offering a scalable pathway to exploit quantum advantages in computational fluid dynamics.

\end{abstract}

\maketitle
\section{\label{sec1.0}Introduction}

Computational fluid dynamics (CFD) is a mature discipline dedicated to simulating fluid flows for both scientific inquiry and engineering practice. Achieving high-precision simulations, however, demands enormous computational resources, which has become a critical bottleneck in contemporary research and industrial applications~\cite{wendt2008computational,slotnick2014cfd,drikakis2019multiscale,trebotich2024exascale,bhatti2020recent}. Conventional CFD predominantly relies on the numerical solution of the Navier–Stokes equations~\cite{constantin1988navier}, whose computational complexity scales with the Reynolds number $R_e$ as $O(R_e^{3})$~\cite{succi2023quantum,nejadmalayeri2013reynolds}, rendering high-$R_e$ simulations prohibitively expensive.

Quantum computing has emerged as a promising avenue for alleviating these computational challenges in CFD~\cite{succi2023quantum}. A variety of quantum algorithms have been devised to accelerate the solution of systems of linear equations~\cite{harrow2009quantum,low2017optimal,gilyen2019quantum}, linear differential equations~\cite{berry2014high,berry2017quantum,xin2020quantum}, and nonlinear differential equations~\cite{liu2021efficient}. These algorithms have been applied to CFD-related problems, including the Poisson equation~\cite{wang2020quantum}, the Burgers equation~\cite{chen2024enabling}, the Korteweg–de Vries equation~\cite{kupershmidt1989quantum}, and acoustic-wave equations~\cite{chen2024enabling}.

Nevertheless, quantum CFD (QCFD) algorithms output quantum states that encode flow field information rather than classical data. Extracting these data requires projective measurements, introducing substantial overhead that can nullify the expected quantum speedups—a fundamental limitation known as the ``output problem''~\cite{aaronson2015read}. For practical flow analyses, reconstructing an entire flow field via quantum measurements is infeasible because the number of required measurements grows exponentially with system dimension.

A practical remedy is to observe that most flow-analysis tasks require only low-dimensional flow features, not the full field. This observation motivates reduced-order modeling, analogous to classical techniques such as Proper Orthogonal Decomposition (POD)~\cite{berkooz1993proper} and Dynamic Mode Decomposition (DMD)~\cite{schmid2010dynamic}. Existing quantum reduced-order approaches can be broadly classified into two families: quantum linear-algebra techniques~\cite{harrow2009quantum,QPCA2014,martyn2021grand} and variational quantum algorithms (VQAs)~\cite{cerezo2021variational,williams2024addressing}.

Quantum linear-algebra methods—such as quantum principal component analysis (QPCA)~\cite{QPCA2014}, quantum support vector machines (QSVMs)~\cite{rebentrost2014quantum,yuan2023quantum,nivelkar2023quantum}, and quantum dynamic mode decomposition (QDMD)~\cite{xue2023quantum}—provide provable guarantees but offer limited versatility across disparate tasks. By contrast, VQAs attain wider applicability through parameterized quantum circuits. Representative examples include multigrid-encoded reduced-order models~\cite{jaksch2023variational}, SPDE-Q-Net for convection–diffusion problems~\cite{yadav2023qpde}, and hybrid quantum–classical neural networks for fluid-dynamics applications~\cite{bazgir2024hybrid,sedykh2024hybrid}. Despite this flexibility, VQAs encounter two pressing issues: the popular hardware-efficient variational quantum circuits (VQCs) lack rigorous expressivity guarantees, and their training is prone to local minima owing to sensitivity to parameter initialization~\cite{liu2023mitigating}.

To overcome these limitations, we introduce an orthogonal-polynomial-based quantum reduced-order model (\OPQROM) for the dimensionality reduction and analysis of quantum-encoded flow field data. Our methodology leverages orthogonal-polynomial basis transformations to construct orthogonal-polynomial quantum neural networks (OPQNNs), thereby improving circuit expressivity while retaining stability. 
Following dimension reduction by the OPQNNs, the resulting compact representations undergo quantum or classical post-processing to execute downstream tasks such as flow field reconstruction and flow-pattern classification. This scheme efficiently captures high-dimensional flow characteristics while retaining the benefits of quantum computation.


We assess \OPQROM on representative CFD benchmarks involving flow field reconstruction and classification. For reconstruction, \OPQROM attains high fidelity across diverse datasets, outperforming classical Chebyshev-based fitting methods in approximating target states. It also exhibits marked improvements in training stability and convergence rate over hardware-efficient VQC-based QNNs, attributable to informed parameter initialization and calibrated circuit design. For classification, \OPQROM achieves higher accuracy than both classical convolutional neural networks (CNNs) and conventional hardware-efficient VQC-based QNNs, while requiring fewer parameters and lower computational effort. These results highlight quantum computing’s promise for efficient analysis of high-dimensional flow fields and verify the utility of orthogonal-polynomial transformations in quantum-network construction. By simultaneously enhancing expressivity and training robustness, our framework delivers new insights and practical guidance for incorporating quantum computing into CFD.

The remainder of the paper is organized as follows. Section~\ref{sec2.0} presents the orthogonal-polynomial quantum neural network in detail. Section~\ref{sec3.0} investigates flow field reconstruction using a quantum linear combination of unitaries (LCU) module and evaluates its fidelity against numerical data. Section~\ref{sec4.0} examines flow field classification by coupling the network with classical fully connected layers and compares its accuracy and efficiency with conventional approaches. Section~\ref{sec5.0} concludes the paper by summarizing the main contributions and outlining future research directions.

\begin{figure*}[!htbp]
    \centering
    \includegraphics[height=10cm,width=17cm]{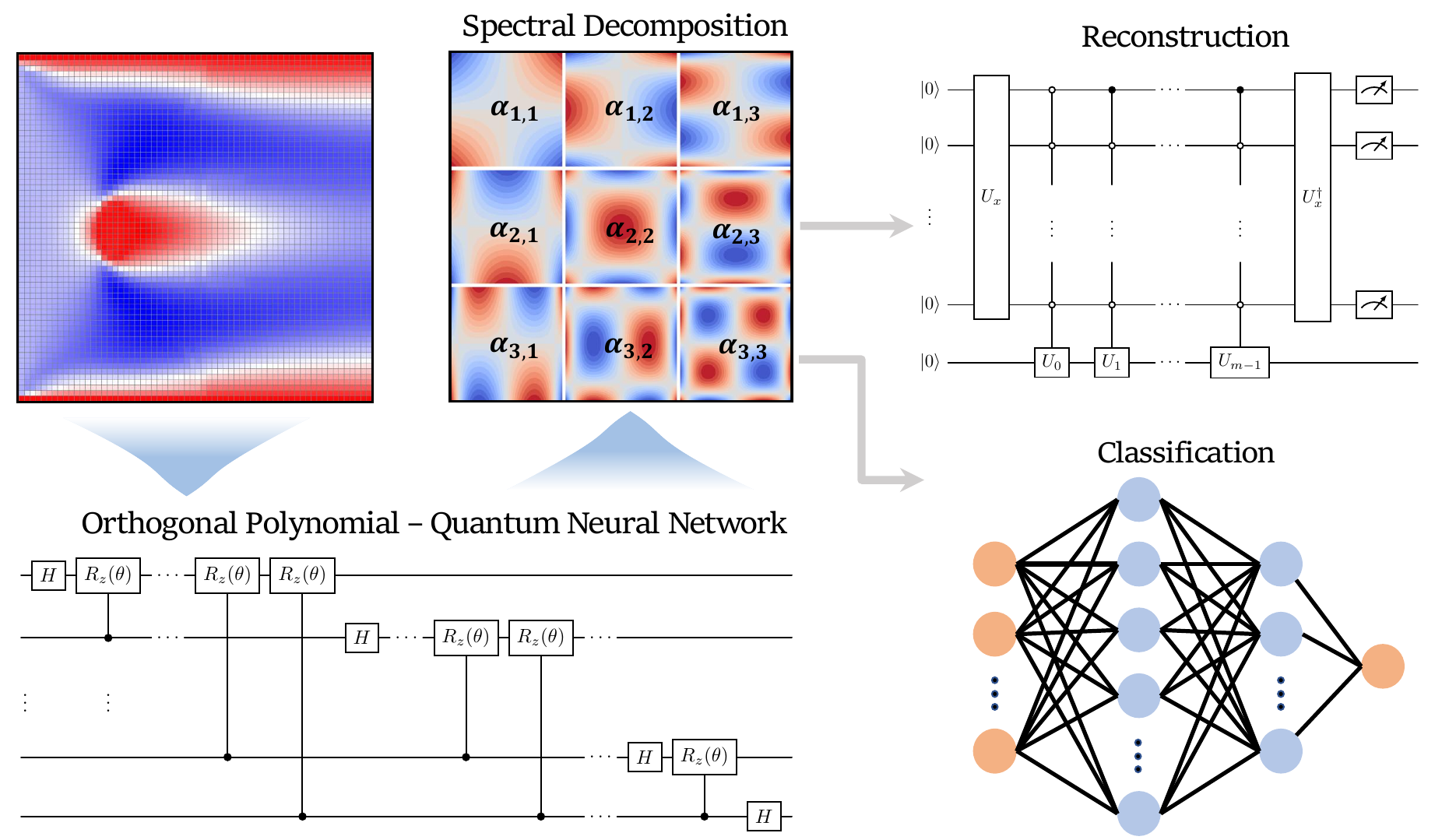}
    \caption{Schematic of \OPQROM applied to flow field reconstruction and classification. The quantum circuit, inspired by orthogonal polynomials, processes the quantum-encoded flow field state. Gates labelled by~$\theta$ are trainable and are optimized to extract low-dimensional features. The resulting features are routed to two downstream modules: a LCU layer for flow field reconstruction and a classical fully connected layer for classification.}
    \label{fig:Schematic}
\end{figure*}

\section{\label{sec2.0}Orthogonal-Polynomial-Based Quantum Neural Network}

The core innovation of \OPQROM lies in harnessing the approximation capabilities of orthogonal polynomials to construct a quantum neural network, which enables dimensionality reduction of quantum-encoded flow field data through parameter optimization. Subsequent quantum or classical post-processing of the compressed representations facilitates specific fluid dynamics tasks. The proposed framework, as illustrated in Fig.~\ref{fig:Schematic}, starts with a quantum state encoding the input flow field data. Then, the state is decomposed into orthogonal polynomial bases via our proposed quantum neural network, named OPQNN, generating expansion coefficients $\alpha$ that capture essential flow features. These coefficients then propagate through task-specific computational workflows where loss functions are evaluated and parameters (including OPQNN components) are iteratively optimized to achieve objectives such as flow reconstruction or pattern classification.

\subsection{\label{sec2.1}Orthogonal-polynomial-based Variational Quantum Circuits}

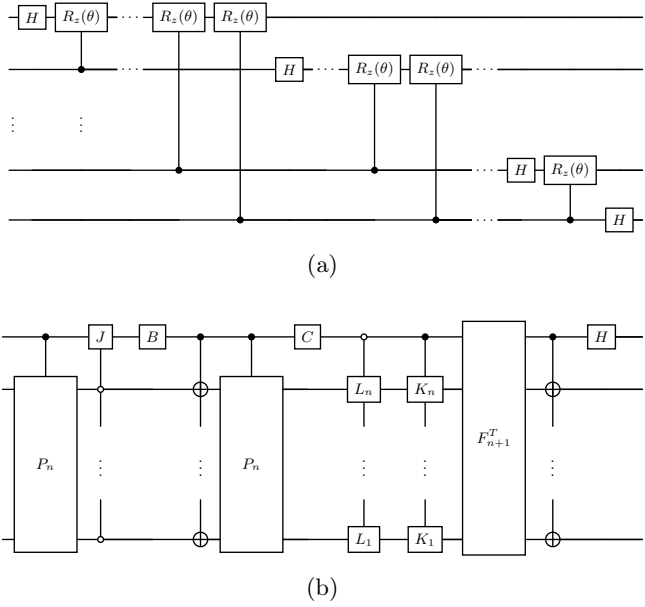
\begin{figure}[!htbp]
    \centering 

     \begin{subfigure}{\linewidth} 
        \centering 
        \resizebox{\linewidth}{!}{%
            \begin{quantikz}[column sep=0.2cm] 
            &\gH &\gRz &\gHdots &\gRz &\gRz & \qw & \qw & \qw & \qw & \qw & \qw & \qw & \qw &\qw \\
            & &\ctrl{-1} &\gHdots & \qw & \qw &\gH &\gHdots &\gRz &\gRz &\gHdots & \qw & \qw & \qw &\qw \\
            \lstick{\makebox[0pt][l]{\hspace{5pt}\vdots}} & \setwiretype{n} & \setwiretype{n}\vdots & \setwiretype{n} & \setwiretype{n} & \setwiretype{n} & \setwiretype{n} & \setwiretype{n} & \setwiretype{n} & \setwiretype{n} & \setwiretype{n} & \setwiretype{n} & \setwiretype{n} & \setwiretype{n} \\
            & & \qw & \qw &\ctrl{-3} & \qw & \qw & \qw &\ctrl{-2} & \qw &\gHdots &\gH &\gRz & \qw &\qw \\
            & & \qw & \qw & \qw &\ctrl{-4} & \qw & \qw & \qw &\ctrl{-3} &\gHdots & \qw &\ctrl{-1} &\gH &\qw
            \end{quantikz}
        }%
        \caption{} 
        \label{fig:combined_circuitsb} 
    \end{subfigure}

    \vspace{\baselineskip} 

   \begin{subfigure}{\linewidth} 
        \centering 
        \resizebox{\linewidth}{!}{%
            \begin{quantikz}[column sep=0.56cm] 
            &\ctrl{1} &\gJ &\gB &\ctrl{2} &\ctrl{1} &\gC &\octrl{2} &\ctrl{2} &\gFtnp &\ctrl{2} &\gH &\qw \\
            &\gPn &\octrl{1}\wire[u]{q} & \qw &\targ{} &\gPn & \qw &\gLn &\gKn & \qw &\targ{} & \qw &\qw \\
            &\setwiretype{n} & \setwiretype{n} & \setwiretype{n} & \setwiretype{n} & \setwiretype{n} & \setwiretype{n}& \setwiretype{n} &\setwiretype{n} & \setwiretype{n} & \setwiretype{n} & \setwiretype{n} &\setwiretype{n} \\
            &\setwiretype{n} &\gdot & \setwiretype{n} &\gdot & \setwiretype{n} & \setwiretype{n} &\gdot &\gdot & \setwiretype{n} &\gdot & \setwiretype{n} &\setwiretype{n} \\
            &\setwiretype{n} & \setwiretype{n} & \setwiretype{n} & \setwiretype{n} & \setwiretype{n} & \setwiretype{n} & \setwiretype{n} & \setwiretype{n} & \setwiretype{n} & \setwiretype{n} & \setwiretype{n} &\setwiretype{n} \\
            & &\octrl{-1} & \qw &\targ{}\wire[u]{q} & \qw & \qw &\gLOne\wire[u]{q} &\gKOne\wire[u]{q} & \qw &\targ{}\wire[u]{q} & \qw &\qw
            \end{quantikz}
        }%
        \caption{ } 
        \label{fig:combined_circuitsa} 
    \end{subfigure}

    \caption{Core variational quantum circuits (VQCs) employed in OPQNN. (a) A QFT-inspired circuit, where each $\theta$ denotes an independent trainable parameter. (b) A QDCT-inspired circuit. Although the trainable parameters are not explicitly drawn, they are embedded within the transformation modules $F_{n+1}$. Detailed circuit descriptions are provided in Appendix~\ref{app:qdct_details}.} 
    \label{fig:combined_circuits} 
\end{figure}

Orthogonal polynomials form mutually orthogonal basis functions under prescribed inner-product conditions~\cite{szeg1939orthogonal}. For dimensionality reduction, projecting target functions onto these bases yields low-dimensional representations that retain salient patterns while eliminating redundancy, thereby enabling efficient feature extraction and compression.

In \OPQROM, we develop orthogonal-polynomial-based quantum neural networks (OPQNNs), capitalizing on well-distributed orthogonal bases to reduce the number of parameters and improve training efficiency. 
Within \OPQROM, we construct orthogonal-polynomial quantum neural networks (OPQNNs) that exploit the uniform distribution of these bases to decrease parameter count and accelerate training. Specifically, OPQNNs embed trainable parameters into orthogonal-polynomial basis transformations; these parameters are optimized via task-driven loss functions. Fig.~\ref{fig:combined_circuits} illustrates parameterized quantum circuits realizing Fourier-basis (quantum fourier transform, QFT~\cite{nielsen2010quantum}) and Chebyshev-basis (quantum discrete cosine transform, QDCT~\cite{klappenecker2001discrete}) transformations. In Fig.~\ref{fig:combined_circuitsb}, each $\theta$ is an independent parameter, whereas in Fig.~\ref{fig:combined_circuitsa} the module $F_{n+1}$ implements the parameterized QDCT.

Because the QDCT employs an ancilla qubit, its successful execution requires measuring that qubit in the $|0\rangle$ state. With random parameter initialization, the post-measurement operation can become non-unitary, yielding a non-orthogonal basis. Hence, for an OPQNN $U_{\theta}$ we have
\begin{equation}\label{eq-u-theta}
U_{\theta}|i\rangle=\alpha_i|\xi_i\rangle|0\rangle+|\psi^{\perp}\rangle,
\end{equation}
where $|\xi_i\rangle$ denotes the $i$-th basis function (Eq.~(\ref{eq-u-theta}) reduces to $U_{\theta}|i\rangle=|\xi_i\rangle$ when no ancilla qubit is present).

\subsection{\label{sec2.2}Dimension Reduction}

After defining the OPQNN circuit, we expand the $n$-dimensional quantum-encoded flow field state $|\psi\rangle$ in the bases $\{|\xi_i\rangle\}$ generated by the network.  
Choosing $m\ll n$ basis states yields
\[
A=[\,\xi_0,\xi_1,\dots,\xi_{m-1}\,],\qquad
\mathbf{x}=[\,x_0,x_1,\dots,x_{m-1}\,]^{\mathsf T},
\]
with
\begin{equation}
\mathbf{x}=\arg\min_{\mathbf{x}}\|A\mathbf{x}-\psi\|.
\end{equation}
The normal-equation solution is
\begin{equation}
\mathbf{x}=(A^{\mathsf T}A)^{-1}A^{\mathsf T}\psi,
\end{equation}
and, with Tikhonov regularization~\cite{golub1999tikhonov},
\begin{equation}
\mathbf{x}=(A^{\mathsf T}A+\lambda I)^{-1}A^{\mathsf T}\psi,
\label{eq-x}
\end{equation}
where $\lambda$ is the regularization parameter and $I$ the identity matrix.  
The matrices $A^{\mathsf T}A$ and $A^{\mathsf T}\psi$ are of sizes $m\times m$ and $m$, respectively; we evaluate every entry via a Hadamard test before solving Eq.~\eqref{eq-x}.

The matrices $A^{\mathsf T}A$ and $A^{\mathsf T}\psi$ are of sizes $m\times m$ and $m$, respectively; we evaluate every entry via a Hadamard test before solving Eq.~\eqref{eq-x}.

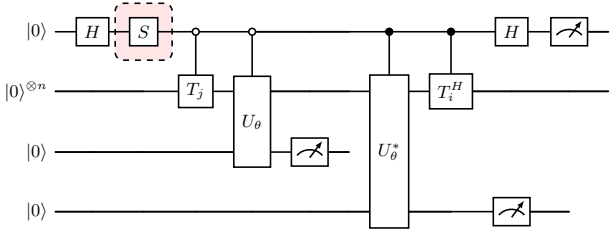
\begin{figure}[htbp]
    \centering
    \begin{tikzpicture}[scale=0.75, transform shape]
    \node[scale=1.0]{
    \begin{quantikz}[column sep=0.37cm]
    \lstick{\ket{0}} &\gH &  \gate{S} \gategroup[wires=1, steps=1, style={dashed, rounded corners, fill=red!10}, label style={above=1ex, text=red}, background]{}&\octrl{1} &\octrl{1} & \qw &\qw &\ctrl{1} &\ctrl{1}  &\gH &\gMeasure{} &\qw \\ 
    \lstick{$\ket{0}^{\otimes n}$} & \qw &\qw&\gTj &\gUtheta & \qw &\qw &\gUthetaStar &\gTiH  & \qw & \qw &\qw \\ 
    \lstick{\ket{0}} & \qw&\qw & \qw & \qw &\gMeasure{} &\qw &\setwiretype{n} &\setwiretype{n} & \setwiretype{n} & \setwiretype{n} &\setwiretype{n} \\ 
    \lstick{\ket{0}} & \qw &\qw& \qw & \qw & \qw &\qw & \qw & \qw &\gMeasure{} & \qw  &\setwiretype{n} 
    \end{quantikz}
    };
    \end{tikzpicture}
    \caption{\centering Hadamard test circuit for estimating inner products.  
    The two bottom qubits are ancillas used in the dual queries of $U_{\theta}$.  
    An $S$ gate is introduced when evaluating imaginary parts.}
    \label{fig:hadamard-test}
\end{figure}

\paragraph*{\textbf{Compute $A^{\mathsf T}A$.}}  
Because $(A^{\mathsf T}A)_{ij}=\langle\xi_i^{*}|\xi_j\rangle$, each inner product is estimated by the Hadamard test (Fig.~\ref{fig:hadamard-test}).  
Here $T_i|0\rangle^{\otimes n}=|i\rangle$ and $U_{\theta}$ is the OPQNN defined in Sec.~\ref{sec2.1}.  
Outcomes are retained only when both ancilla qubits are measured in $|0\rangle$.

\paragraph*{\textbf{Compute $A^{\mathsf T}|\psi\rangle$.}}  
The $i$-th component, $\langle\xi_i^{*}|\psi\rangle$, is obtained with a Hadamard test that queries $U_{\theta}$ and the state-preparation routine for $|\psi\rangle$.

\paragraph*{\textbf{Solve for $\mathbf{x}$.}}  
After estimating $A^{\mathsf T}A$ and $A^{\mathsf T}\psi$, Eq.~\eqref{eq-x} yields $\mathbf{x}$.

\subsection{\label{sec2.3}Post-Processing}

With the expansion coefficients $\mathbf{x}$ in hand, \OPQROM supports diverse post-processing architectures tailored to specific tasks.  
For flow field reconstruction, the orthogonal bases together with least-squares coefficients are mapped back to a high-dimensional quantum state via an LCU layer, and training minimizes quantum-state fidelity loss.  
For flow-pattern classification, the low-dimensional feature vector $\mathbf{x}$ is passed to classical fully connected layers.  
Experiments confirm the robustness of \OPQROM in both reconstruction and classification, underscoring its versatility for high-dimensional data analysis.  
Subsequent sections elaborate on implementation details, including network architecture, hardware considerations, and gradient-based training.

For a task-specific loss $\mathcal{L}$, gradients with respect to parameters $\theta_i$ are approximated via finite differences,
\begin{equation}
\frac{\partial\mathcal{L}}{\partial\theta_i}\approx
\frac{\mathcal{L}(\theta_i+\epsilon)-\mathcal{L}(\theta_i-\epsilon)}{2\epsilon},
\end{equation}
after which optimizers such as Adam update the parameters.

\section{\label{sec3.0}Flow Field Reconstruction}



We first examine flow field reconstruction.  After the network yields the fitting coefficients \(\mathbf{x}\), these values are combined with the orthogonal basis states \(\{|\xi_i\rangle\}_{i=0}^{m-1}\).  Using the LCU algorithm, the coefficients are applied to their corresponding basis states to reconstruct an approximate target state
\begin{equation}
|\tilde{\psi}\rangle=\sum_{i=0}^{m-1}x_i|\xi_i\rangle.
\end{equation}

\begin{figure}[h]
    \centering
    \begin{tikzpicture}[scale=0.9, transform shape]
    \node[scale=1.0]{
    \begin{quantikz}
    \lstick{\ket{0}} &\gate[wires=6]{U_{x}} &\octrl{1} &\ctrl{1} &\gHdots &\ctrl{1} &\gate[wires=6]{U_{x}^{\dagger}} &\meter{} &\qw \\ 
    \lstick{\ket{0}} & \qw &\octrl{1} &\octrl{1} &\gHdots &\octrl{1} & \qw &\meter{} &\qw \\ 
      &\setwiretype{n}  & \setwiretype{n} &\setwiretype{n} &\setwiretype{n} & \setwiretype{n} & \setwiretype{n} & \setwiretype{n}& \setwiretype{n} \\ 
    \lstick{\hspace{-1.8em}\raisebox{1.3em}\vdots} &\setwiretype{n} &\vdots &\vdots & \setwiretype{n} &\vdots & \setwiretype{n} &\setwiretype{n} & \setwiretype{n} \\ 
      &\setwiretype{n} & \setwiretype{n} & \setwiretype{n} & \setwiretype{n} & \setwiretype{n} & \setwiretype{n} & \setwiretype{n} & \setwiretype{n} \\ 
    \lstick{\ket{0}} & \qw &\octrl{1}\wire[u]{q} &\octrl{1}\wire[u]{q} &\gHdots &\octrl{1}\wire[u]{q} & \qw &\meter{} &\qw \\ 
    \lstick{\ket{0}} & \qw &\gate{U_0} &\gate{U_1} &\gHdots &\gate{U_{m-1}} & \qw & \qw & \qw 
    \end{quantikz}
    
    };
    \end{tikzpicture}
    \caption{\centering LCU quantum circuit.}
    \label{fig:LCU circuit}
\end{figure}
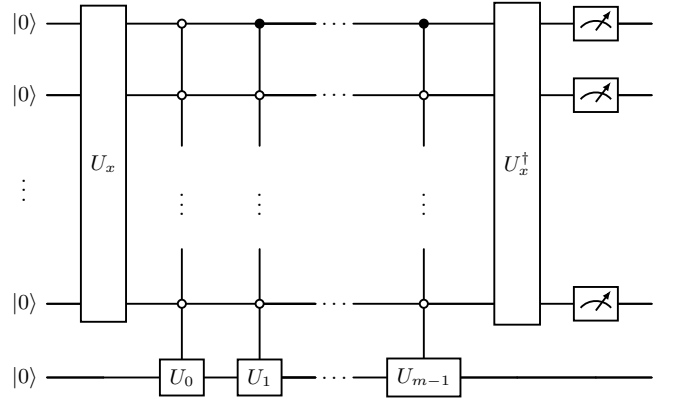

The circuit in Fig.~\ref{fig:LCU circuit} implements this procedure, where \(U_x\) is an amplitude-encoding circuit that loads the coefficients \(\{x_1,x_2,\dots,x_m\}\), and each \(U_i\) prepares the basis state \(|\xi_i\rangle\) for \(i=0,1,\dots,m-1\).

\begin{figure}[h]
    \centering
    \begin{tikzpicture}[scale=1.0, transform shape]
    \node[scale=1.0]{
    \begin{quantikz}
    \lstick{$|0\rangle$} &\gate{H} &\ctrl{1} &\gate{H} &\meter{} &\qw \\ 
    \lstick{$|\psi\rangle$} &\qw &\gate[wires=2]{\text{SWAP}} &\qw &\qw &\qw \\ 
    \lstick{$|\widetilde{\psi}\rangle$} &\qw & \qw &\qw &\qw &\qw 
    \end{quantikz}
    
    };
    \end{tikzpicture}
    \caption{\centering SWAP test circuit for fidelity estimation.}
    \label{fig:SWAP test}
\end{figure}
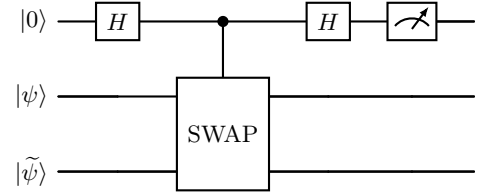

Reconstruction quality is quantified by the fidelity
\begin{equation}
F = |\langle\psi|\tilde{\psi}\rangle|^{2}, \qquad 0\le F\le1,
\end{equation}
with \(F=1\) indicating perfect agreement.  Fidelity is evaluated via the SWAP test (Fig.~\ref{fig:SWAP test}), where measuring the ancilla in \(|0\rangle\) with probability \(P(0)\) yields \(F=2P(0)-1\).  During training we minimise \(1-F\) as the loss.


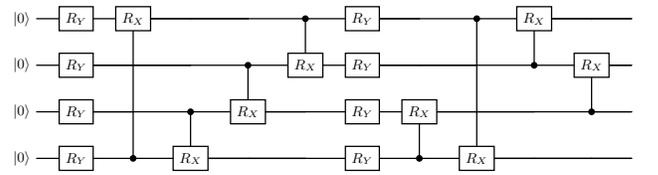
\begin{figure}[h]
    \centering
    \begin{tikzpicture}[scale=0.6, transform shape]
    \node[scale=1.0]{
    \begin{quantikz}
    \lstick{\ket{0}} &\gate{R_Y} &\gate{R_X} & \qw & \qw &\ctrl{1} &\gate{R_Y} & \qw &\ctrl{3} &\gate{R_X} & \qw& \qw \\ 
    \lstick{\ket{0}} &\gate{R_Y} & \qw & \qw &\ctrl{1} &\gate{R_X} &\gate{R_Y} & \qw & \qw &\ctrl{-1} &\gate{R_X}& \qw \\ 
    \lstick{\ket{0}} &\gate{R_Y} & \qw &\ctrl{1} &\gate{R_X} & \qw &\gate{R_Y} &\gate{R_X} & \qw & \qw &\ctrl{-1}& \qw \\ 
    \lstick{\ket{0}} &\gate{R_Y} &\ctrl{-3} &\gate{R_X} & \qw & \qw &\gate{R_Y} &\ctrl{-1} &\gate{R_X} & \qw & \qw & \qw 
    \end{quantikz}
    
    };
    \end{tikzpicture}
    \caption{\centering Hardware-efficient ansatz.}
    \label{fig:hardware-efficient ansatz}
\end{figure}

\begin{figure*}[ht]
    \centering
    \includegraphics[width=0.9\textwidth]{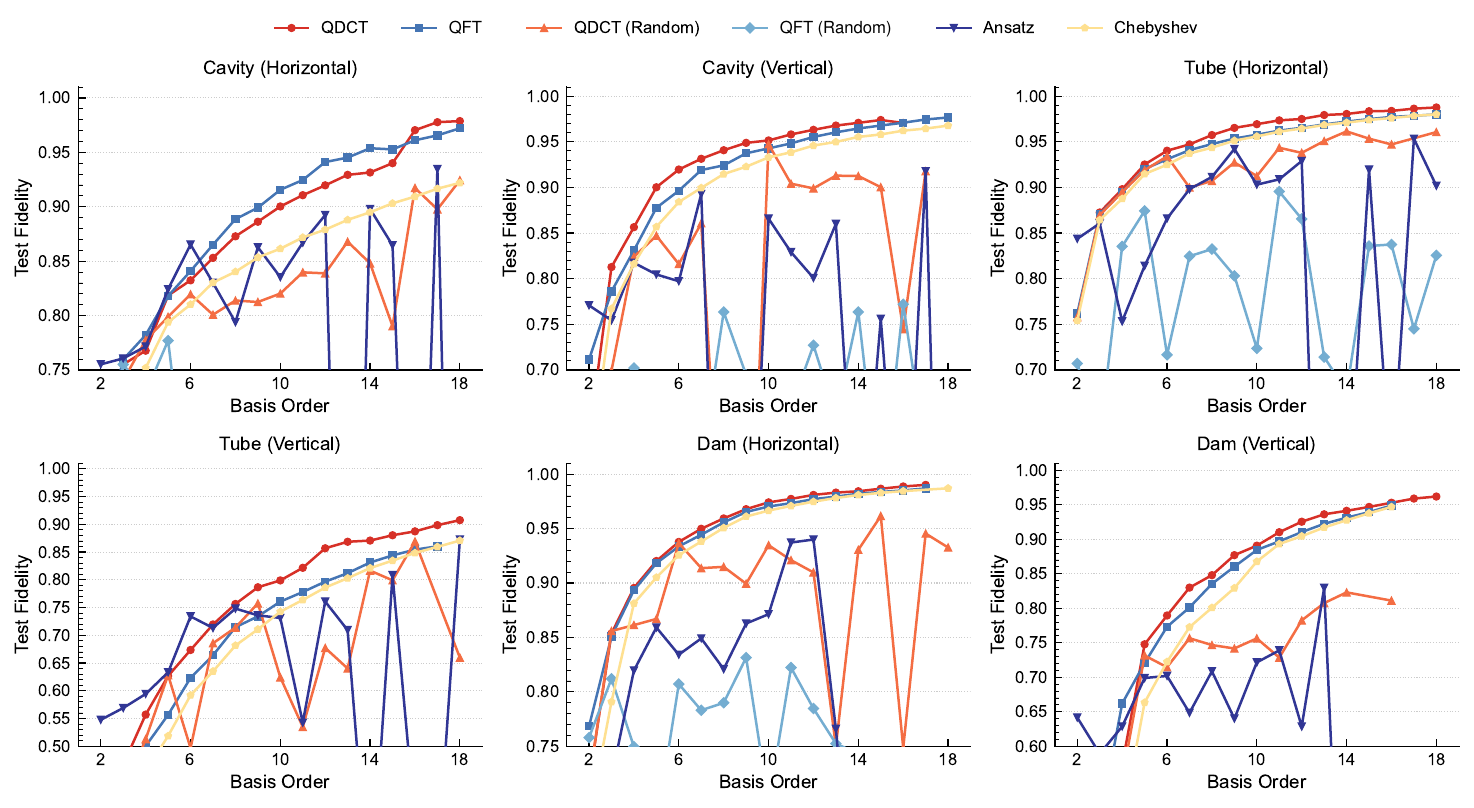}
    \caption{Test-set fidelity as a function of reconstruction order \(m\) (i.e., the number of orthogonal bases). Curves denote different models: red, QDCT-based; dark blue, QFT-based; orange-red and medium blue, their randomly initialised counterparts; deep blue, hardware-efficient ansatz; yellow, classical Chebyshev fitting. Across all datasets, QDCT- and QFT-based networks achieve the highest fidelities and exhibit superior generalization and numerical stability relative to the classical baseline.}
    \label{fig:Fidelity_results}
\end{figure*}

\subsection{\label{sec3.1}Setup}

Our numerical study validates the proposed reconstruction algorithm through dataset choice, parameter settings, baselines, metrics, and analysis.

\textbf{Dataset.}  We employ \textsc{CFDBench}~\cite{luo2023cfdbench}, a benchmark containing four canonical flows—cavity, tube, dam, and cylinder—under varying boundary conditions (BC), fluid properties (PROP), and geometries (GEO).  All simulations are interpolated to a \(64\times64\) grid holding horizontal and vertical velocity components.

\textbf{Partitioning.}
\begin{itemize}
    \item \emph{Minimal-class partition}: for a single flow type, fix one operating condition and use either \(u\) or \(v\) velocities across time.  The resulting samples are split into training and testing sets.
    \item \emph{Comprehensive partition}: for a given flow type, merge all operating conditions, then split into training and testing sets to assess generalisation.
\end{itemize}

\textbf{Quantum implementation.}  Each experiment uses 14 qubits (two registers of 7) to compress a length-64 dimension.  Circuits are built in \texttt{pyQPanda}~\cite{dou2022qpanda} and \texttt{PennyLane}~\cite{bergholm2018pennylane} with  
a learning rate of 0.001 (decayed to 10 \% every 25 epochs), 75 epochs in total, and reconstruction order \(m\) ranging from 2 to 18.

\textbf{Metrics.}  Performance is reported using mean-squared error (MSE) and fidelity.

\subsection{\label{sec3.2}Simulation Results and Analysis}


We conduct three comparative studies:

\paragraph{Classical Chebyshev fitting.}  
Chebyshev polynomials provide strong polynomial approximations.  We compare QDCT- and QFT-based networks against Chebyshev fitting on \(u\) and \(v\) for cavity, tube, and dam flows.

\paragraph{Initialisation strategies.}  
To probe the barren-plateau issue, we contrast our structured \(\pi\) initialisation with random initialisation.

\paragraph{Hardware-efficient ansatz.}  
We benchmark against the expressive hardware-efficient circuit of Fig.~\ref{fig:hardware-efficient ansatz}~\cite{sim2019expressibility}.

\begin{figure*}[!htbp]
    \centering
    \includegraphics[width=0.9\textwidth]{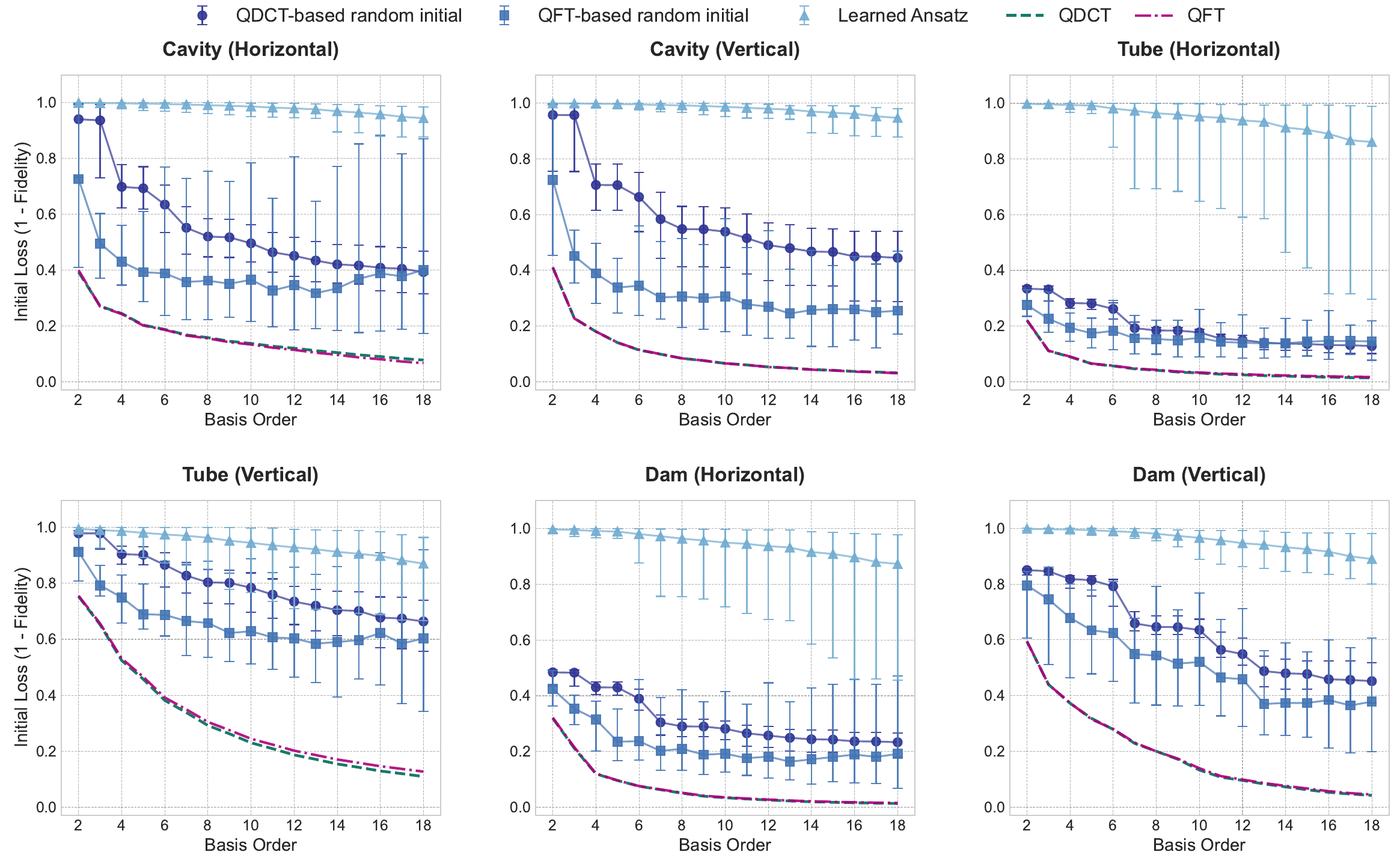}
    \caption{Initial loss \((1-\text{Fidelity})\) versus reconstruction order \(m\) under different initialisation schemes across multiple datasets.  
    Lines represent the proposed QDCT/QFT initialisations; markers with error bars show the means \(\pm\) one standard deviation for randomly initialised QDCT/QFT circuits and for the hardware-efficient ansatz.  
    The structured initialisation consistently yields the lowest initial loss—providing a superior starting point for optimisation compared with random initialisation and the ansatz approach.}
    \label{fig:initial-loss}
\end{figure*}

As illustrated in Fig.~\ref{fig:Fidelity_results}, the six sub-plots correspond to three flow scenarios, each analysed for both horizontal and vertical velocity components. In theory, fidelity should rise with increasing reconstruction order, and the results confirm this trend. Across all experiments, the QDCT- and QFT-based networks consistently attain the highest reconstruction fidelities. Owing to space limitations, detailed results for specific flow cases, boundary conditions, and velocity directions are provided in the appendix.

Relative to classical Chebyshev fitting, both quantum-polynomial approaches deliver uniformly superior fidelity, most notably for the velocity reconstructions of the three representative flows. Among the six datasets, the QDCT-based network generally performs best, indicating that orthogonal-polynomial QNNs not only reduce reconstruction error but also generalise more effectively than classical polynomial methods.

We next examine the impact of different initialisation strategies (Fig.~\ref{fig:initial-loss}). The figure plots the initial loss (defined as $1{-}$Fidelity at the first training step) versus reconstruction order for five configurations: QDCT- and QFT-based circuits with our structured initialisation, their randomly initialised counterparts, and the hardware-efficient ansatz. For the latter three, error bars show the mean and standard deviation over multiple runs, mitigating random-seed effects.

The structured initialisation yields the lowest initial loss at every reconstruction order, outperforming both random initialisation and the ansatz. This demonstrates that a principled starting point appreciably improves optimisation, helping to alleviate barren-plateau issues. Because the QDCT and QFT circuits have mathematically interpretable layouts, their parameters can be meaningfully initialised rather than assigned ad-hoc random values—enhancing both theoretical interpretability and practical training efficiency.

Finally, when compared with the hardware-efficient ansatz and randomly initialised circuits, the QDCT- and QFT-based models achieve higher and more stable fidelities on most flow datasets. At larger reconstruction orders, the ansatz often displays pronounced accuracy fluctuations—likely symptomatic of barren plateaus—whereas the orthogonal-polynomial networks maintain steady, high-quality reconstructions. Given that accurate flow field reconstruction must balance reconstruction order and fidelity, unstable methods offer limited practical value. These findings reaffirm the superior expressiveness and training stability of our orthogonal-polynomial QNNs, which avoid the key optimisation bottlenecks typical of generic parameterised quantum circuits.

\begin{figure}[!htbp]
    \centering

    \begin{subfigure}{0pt}\phantomsubcaption\label{fig9a}\end{subfigure}
    \begin{subfigure}{0pt}\phantomsubcaption\label{fig9b}\end{subfigure}
    \begin{subfigure}{0pt}\phantomsubcaption\label{fig9c}\end{subfigure}
    \begin{subfigure}{0pt}\phantomsubcaption\label{fig9d}\end{subfigure}
    \begin{subfigure}{0pt}\phantomsubcaption\label{fig9e}\end{subfigure}
    \begin{subfigure}{0pt}\phantomsubcaption\label{fig9f}\end{subfigure}

    \includegraphics[width=0.95\linewidth]{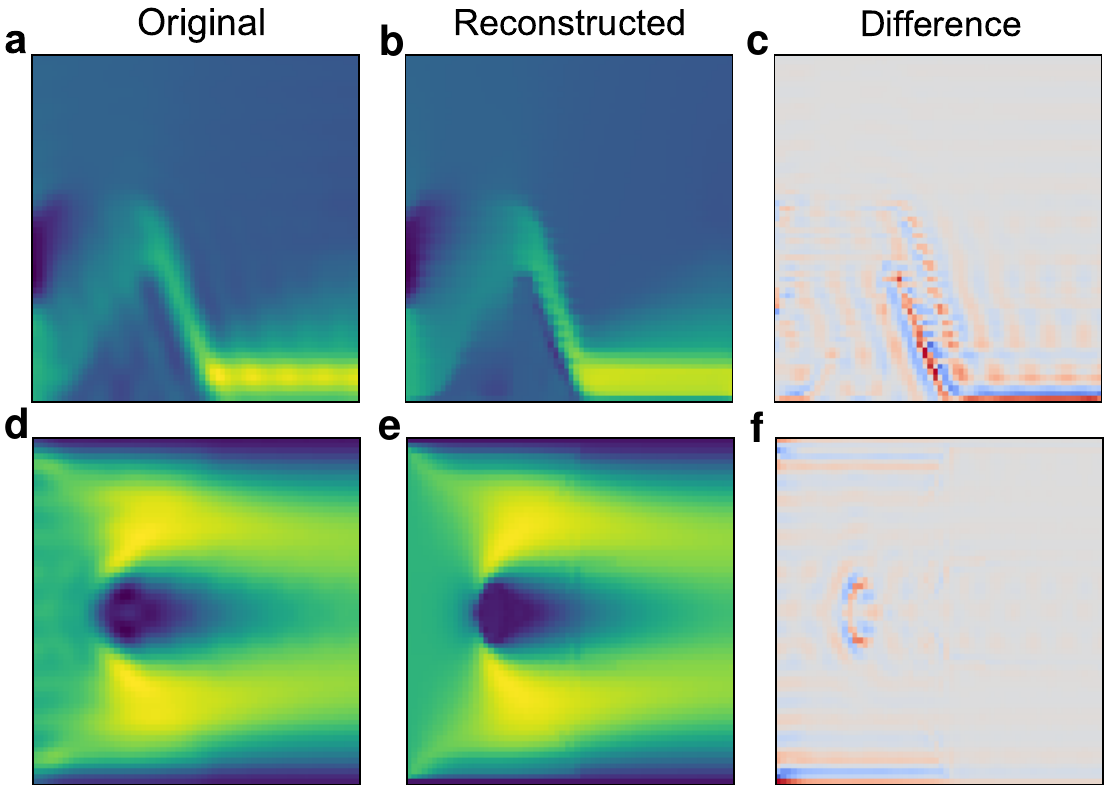}

    \caption{\label{fig:reconstruction_example}Comparison of original and reconstructed flow fields using \OPQROM\ at reconstruction order 8. The subplots within the image display: (a) Original dam break flow, (b) Reconstructed dam break flow, (c) Difference for dam break flow, (d) Original cylinder flow, (e) Reconstructed cylinder flow, (f) Difference for cylinder flow.}
\end{figure}

Fig.~\ref{fig:reconstruction_example} visually demonstrates these capabilities for reconstruction order~8. Fig.~\ref{fig9a} and~\ref{fig9d} present the original dam-break and cylinder-flow fields, respectively; Fig.~\ref{fig9b} and~\ref{fig9e} show their reconstructions; and Fig.~\ref{fig9c} and~\ref{fig9f} plot the point-wise differences. Even at this moderate order, the reconstructions capture essential flow features with high fidelity. Although small discrepancies are visible in the difference maps, the key information is preserved, striking an effective balance between accuracy and circuit depth.

\section{\label{sec4.0}Flow Field Data Classification}

Flow field classification is a canonical analysis task in fluid mechanics; by labelling the dominant structures in simulated or experimental data, it supports systematic evaluation of complex flows in CFD studies, aerodynamic design, and numerous industrial fluid-engineering problems \cite{brunton2020machine,fukami2020assessment}.

\OPQROM extends naturally to flow field classification.  
After obtaining the fitting coefficients \(\mathbf{x}\), a single classical fully connected layer suffices to complete the task, yielding more efficient classification than standard neural networks.  
The quantum layer undertakes feature extraction and dimensionality reduction via a procedure analogous to classical orthogonal-polynomial fitting, while leveraging the potential exponential speed-up of quantum computation to project high-dimensional flow data into a compact feature space.  
This step captures the salient information while sharply reducing both parameter count and computational cost; the resulting low-dimensional features are then passed to the fully connected layer for classification.

To assess performance, we compare our approach with two baselines.  
The first is a hardware-efficient ansatz that performs well on diverse tasks; we adopt a top-performing variant as the quantum benchmark.  
The second benchmark is a classical CNN chosen for its relatively low parameter count, providing a fair classical reference.



\subsection{\label{sec4.1}Numerical Results}




We employ \(64\times64\) flow field data comprising four archetypes—cavity, tube, cylinder, and dam—constituting a four-class problem.  
We compare the hybrid quantum–classical networks (QDCT + FC, QFT + FC) with the hardware-efficient ansatz and the classical CNN, focusing on parameter count, classification accuracy, and computational complexity.

Table~\ref{tab:classification_results} summarises the results.  
With equivalent parameter budgets, QFT + FC attains 99.5 \% accuracy using only 46 parameters, surpassing the CNN’s 98 \% accuracy at the same scale.  
The hardware-efficient ansatz reaches a best accuracy of 98.13 \%, further underscoring the advantage of orthogonal-polynomial quantum networks for flow field classification.

Regarding computational complexity, the CNN must perform multiple convolutions over the full \(64\times64\) grid, incurring \(\mathcal{O}(N^{2}d)\) cost, where \(N^{2}\) is the data dimension and \(d\) the convolution depth (64 in this study).  
This quadratic growth becomes prohibitive for high-resolution problems.  
By contrast, the QDCT- and QFT-based models exploit quantum superposition to complete dimensionality reduction and feature extraction in \(\mathcal{O}(m\log N)\) time, dramatically lowering resource requirements while maintaining expressivity and stable training.

\begin{table}[H]
    \caption{\label{tab:classification_results}Parameter counts, classification accuracies, and complexities for different models.}
    \centering
    \begin{ruledtabular}
    \begin{tabular}{lccc}
        \textbf{Model} & \textbf{Parameters} & \textbf{Accuracy} & \textbf{Complexity} \\
        \hline
        QDCT + FC & 61 & 95.15\% & \textbf{\(O(m \log(N))\)} \\
        QFT + FC & \textbf{46}  & \textbf{99.5\%} & \textbf{\(O(m \log(N))\)} \\
        Ansatz + FC & 61 & 98.13\% & \textbf{\(O(m \log(N))\)} \\
        Best CNN & \textbf{46} & 98\% & \(O(N^2)\) \\
    \end{tabular}
    \end{ruledtabular}
\end{table}

Overall, the proposed hybrid quantum–classical architecture not only surpasses both the CNN and the hardware-efficient ansatz in accuracy but also offers exponentially lower computational complexity, thereby substantially reducing resource demand.  
These results further validate the promise of orthogonal-polynomial quantum networks for flow field classification tasks.


\section{\label{sec5.0}Conclusion}

In this work we introduce \OPQROM, an orthogonal-polynomial quantum reduced-order model that merges orthogonal-polynomial basis transformations with variational quantum circuits (VQCs).  
This synergy increases representational capacity and trainability while compressing quantum-encoded flow field data into compact, low-dimensional representations.  
Numerical experiments demonstrate dual strengths—high-fidelity flow field reconstruction and accurate flow-pattern classification—surpassing classical methods and existing quantum baselines while markedly lowering computational complexity and parameter overhead.  

For flow field reconstruction, \OPQROM employs a new quantum-network architecture that attains exceptional fidelity across diverse datasets, outperforming classical Chebyshev-based approximations by converging more tightly to target flow states.  
A systematic parameter-initialization scheme further stabilizes training and accelerates convergence, underscoring the framework’s scalability and practical feasibility.  
In classification tasks, the quantum model surpasses classical convolutional neural networks (CNNs) in accuracy while reducing parameter counts by orders of magnitude, effectively shifting computational complexity from \(O(N^{2})\) to \(O(m\log N)\).  
These results highlight quantum computing’s transformative potential for high-dimensional fluid-dynamics analysis. 

Despite these advances, two key challenges remain.  
(1) Noisy intermediate-scale quantum (NISQ) devices are limited by qubit count and noise, constraining large-scale flow field simulations.  
Upcoming work will refine the circuit architecture to demonstrate concrete flow-analysis tasks on current NISQ hardware.  
(2) The present implementation employs a single class of basis functions, limiting \OPQROM’s analytical versatility.  
We will extend the framework by integrating multiple, complementary bases, enabling more accurate and efficient analyses of complex flow scenarios.  

This dual pathway—hardware-oriented circuit optimization and basis-diversity enhancement—aims to advance \OPQROM’s practicality for quantum-accelerated CFD and broaden quantum computing applications in complex flow field analysis.

\acknowledgments
This work has been supported by the National Key Research and Development Program of China (Grant No. 2023YFB4502500), the National Natural Science Foundation of China (Grant No. 12404564), and the Anhui Province Science and Technology Innovation (Grant No. 202423s06050001).



\appendix

\section{\label{app:qdct_details}Quantum Discrete Cosine Transform}





This appendix provides the definitions of the fundamental quantum gates comprising the QDCT circuit variant referenced in~\cite{klappenecker2001discrete}, from which our Chebyshev circuit is derived. This QDCT circuit operates on \(n\) qubits, corresponding to a state space of dimension \(N=2^n\).

A key parameter appearing in the circuit's phase gates is \(\omega\), a root of unity defined as:
\begin{equation}
\omega = e^{\frac{\pi i}{2N}} = \cos\left(\frac{\pi}{2N}\right) + i \sin\left(\frac{\pi}{2N}\right),
\end{equation}
where \(N=2^n\). Its complex conjugate is \(\bar{\omega} = e^{-\frac{\pi i}{2N}}\).

The specific gates used in the construction of the underlying QDCT circuit are as follows:

\begin{itemize}
    \item \textbf{Diagonal Phase Gates \(K_j\) and \(L_j\):} These gates introduce phase shifts conditioned on the computational basis, typically with the index \(j\) ranging from 1 to \(n\). They are defined as:
    \begin{equation}
    K_j = \begin{pmatrix} \bar{\omega}^{2^{j-1}} & 0 \\ 0 & 1 \end{pmatrix},
    \end{equation}
    \begin{equation}
    L_j = \begin{pmatrix} 1 & 0 \\ 0 & \omega^{2^{j-1}} \end{pmatrix}.
    \end{equation}
    Up to a global phase, these gates are equivalent to \(R_z\) rotations. Notably, they are diagonal matrices and thus satisfy \(K_j^T = K_j\), \(L_j^T = L_j\).
    
    \item \textbf{Phase Gate \(C\):} A single-qubit, \(R_z\)-type phase gate, defined as:
    \begin{equation}
    C = \begin{pmatrix} 1 & 0 \\ 0 & \bar{\omega} \end{pmatrix} = \begin{pmatrix} 1 & 0 \\ 0 & e^{-\frac{\pi i}{2N}} \end{pmatrix}.
    \end{equation}
    This gate corresponds to \(R_z(-\frac{\pi}{N})\) up to a global phase and is also diagonal (\(C^T = C\)).
    
    \item \textbf{Unitary Gate \(B\):} A single-qubit unitary defined as:
    \begin{equation}
    B = \frac{1}{\sqrt{2}} \begin{pmatrix} 1 & i \\ 1 & -i \end{pmatrix},
    \end{equation}
    which can be constructed, for instance, as \(B = HS\) (\(H\): Hadamard, \(S\): Phase gate). In the Chebyshev circuit, its transpose is employed:
    \[
    B^T = \frac{1}{\sqrt{2}} \begin{pmatrix} 1 & 1 \\ i & -i \end{pmatrix}.
    \]
    
    \item \textbf{Unitary Gate \(J\):} Another single-qubit unitary defined by:
    \begin{equation}
    J = \frac{1}{\sqrt{2}} \begin{pmatrix} 1 & -i \\ -i & 1 \end{pmatrix},
    \end{equation}
    which is also symmetric and thus satisfies \(J^T = J\).
    
    \item \textbf{Quantum Incrementer \(P_n\):} A multi-qubit modular addition gate, performing:
    \begin{equation}
    P_n : |x\rangle \mapsto |(x+1) \bmod N\rangle,
    \end{equation}
    for \(|x\rangle\) denoting the basis state for \(x \in \{0, \dots, N-1\}\). Its implementation relies on controlled logic gates. (As noted in the main text, \((P_n^{-1})^T = P_n\). The QDCT may employ either \(P_n\) or \(P_n^{-1}\), and the Chebyshev circuit uses the appropriately transposed form.)

    \item \textbf{Other Gates:} Standard quantum gates such as the Hadamard (\(H\)) and blocks representing the Quantum Fourier Transform (\(F_{n+1}\)) or its inverse/transpose are also employed, all with their conventional definitions. Importantly, \(H^T = H\). The \(F_{n+1}^T\) block designates the transpose of the standard QFT.
\end{itemize}

\begin{figure*}[htbp]
    \centering
    \includegraphics[width=0.9\textwidth]{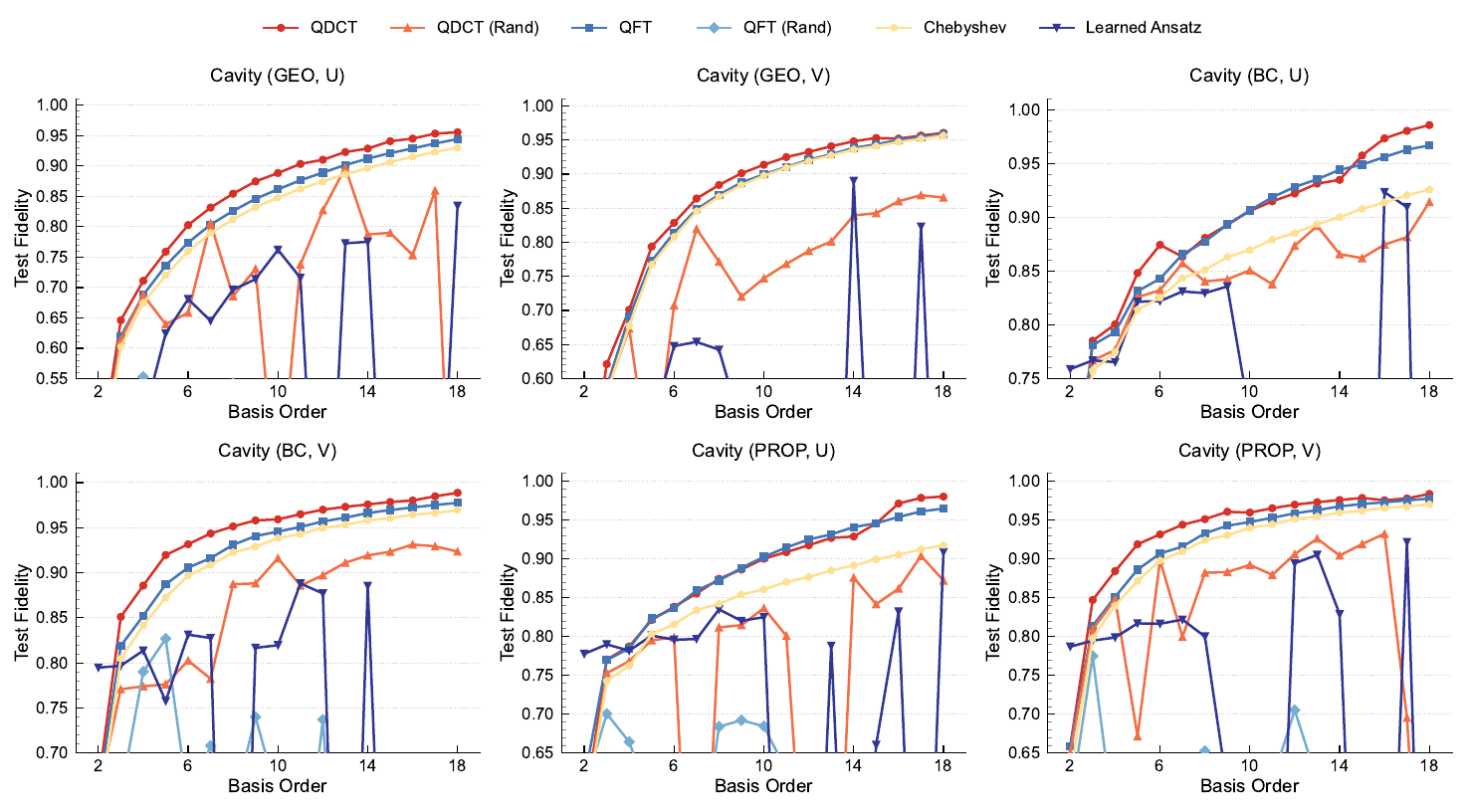}
    \caption{\centering\label{fig:cavity_detailed}Test-set fidelity results for the \texttt{cavity} flow field across all six conditions.}
\end{figure*}

\begin{figure*}[htbp]
    \centering
    \includegraphics[width=0.9\textwidth]{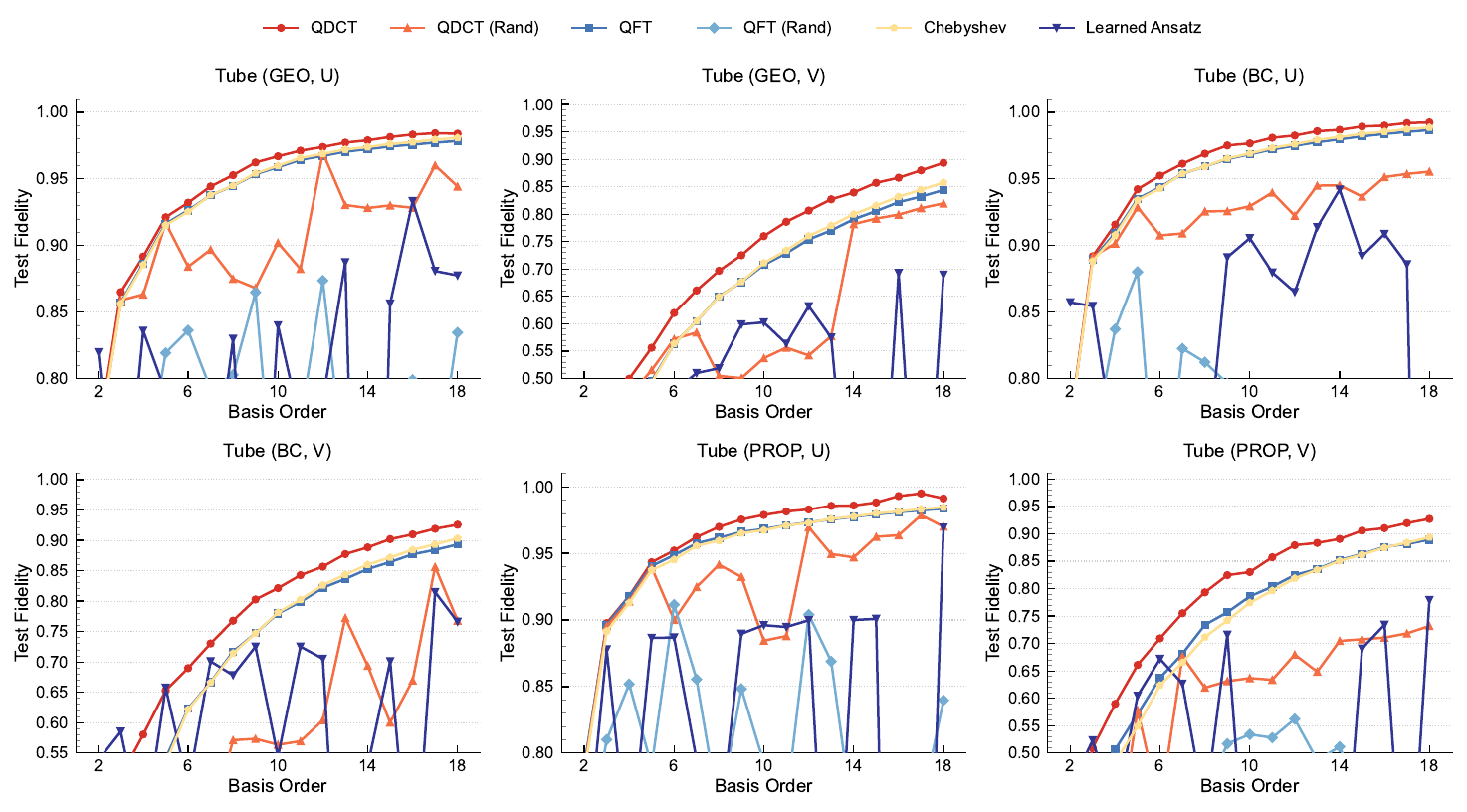}
    \caption{\centering\label{fig:tube_detailed}Test-set fidelity results for the \texttt{tube} flow field.}
\end{figure*}

\begin{figure*}[htbp]
    \centering
    \includegraphics[width=0.9\textwidth]{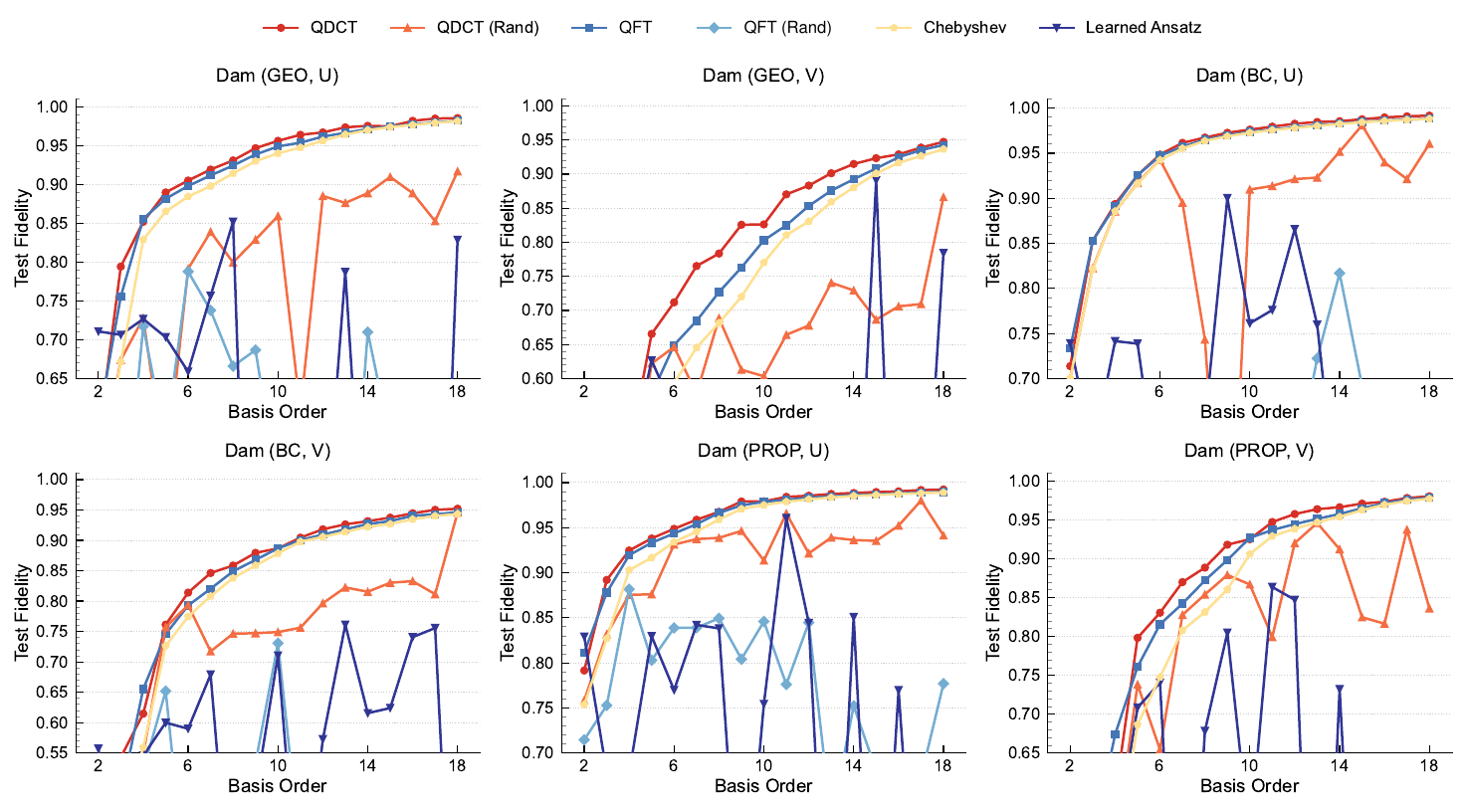}
    \caption{\centering\label{fig:dam_detailed}Test-set fidelity results for the \texttt{dam} flow field.}
\end{figure*}





\section{\label{formres}Detailed Fidelity Results for All Flow Conditions and Directions}

In this appendix, we focus on the \textbf{Minimal-class partition} data partition strategy. Specifically, for a single flow type (e.g., Cavity Flow), a specific operating condition (a particular \texttt{geo} or \texttt{bc}) is selected, and either the horizontal or vertical velocity component is extracted. The data for each time step is then unfolded, merged, and split into training and test sets for evaluating the quantum reconstruction performance. This partition strategy differs from the one used in the main text, where the \textbf{Comprehensive partition} strategy is employed.

This appendix presents detailed reconstruction fidelity results for all 18 datasets across the three flow field scenarios (\texttt{cavity}, \texttt{tube}, and \texttt{dam}), evaluated under three conditions: geometry variation (geo), boundary condition variation (bc), and property variation (prop), with both horizontal ($u$) and vertical ($v$) velocity components. The following figures provide expanded results from those summarized in the main text.

Each subfigure shows the test-set fidelity versus the reconstruction order $m$ for six methods:
\begin{itemize}
    \item \textbf{QDCT (Red):} The proposed method based on quantum discrete cosine transform. Represented by the key \texttt{small\_qdct}.
    \item \textbf{QFT (Dark Blue):} A quantum discrete Fourier transform variant. Represented by the key \texttt{small\_qft}.
    \item \textbf{QDCT (Random) (Orange-Red):} QDCT with randomly initialized encoding. Represented by the key \texttt{small\_qdct\_random}.
    \item \textbf{QFT (Random) (Medium Blue):} QFT with randomly initialized encoding. Represented by the key \texttt{small\_qft\_random}.
    \item \textbf{Learned Ansatz (Deep Blue):} Hardware-efficient variational quantum circuit (VQC), using the learned parameters. Represented by the key \texttt{small\_new\_anstaz}. 
    \item \textbf{Chebyshev (Yellow):} Classical reconstruction using Chebyshev basis fitting. Represented by the key \texttt{chebyshev\_basis}.
\end{itemize}

Note that for clarity, the lower limit of the vertical axis (Test Fidelity) is adjusted individually for each subfigure based on the data range, while the upper limit is consistently set near 1.0.

The detailed fidelity trends for each of the three primary flow field types are presented sequentially. Fig.~\ref{fig:cavity_detailed} shows the results for the \texttt{cavity} flow field, Fig.~\ref{fig:tube_detailed} presents the results for the \texttt{tube} flow field, and Fig.~\ref{fig:dam_detailed} displays the results for the \texttt{dam} flow field.

\paragraph*{\textbf{Summary Observations:}} Across all datasets and reconstruction orders, the QDCT and QFT methods consistently demonstrate superior fidelity compared to classical approaches. Their performance improves steadily with increasing reconstruction order, and structured initialization proves significantly more stable than random variants, which tend to exhibit higher variance and early saturation. The hardware-efficient ansatz performs poorly in high-dimensional reconstructions, likely due to barren plateau effects that hinder optimization. In contrast, the classical Chebyshev method reaches lower fidelity levels and saturates early, reflecting its limited expressive capacity in complex, high-dimensional settings. These results collectively confirm the robustness and generalization ability of QDCT-based and QFT-based quantum neural networks for flow field reconstruction.

\bibliography{polyqrom.bib}

\begin{thebibliography}{44}%
\makeatletter
\providecommand \@ifxundefined [1]{%
 \@ifx{#1\undefined}
}%
\providecommand \@ifnum [1]{%
 \ifnum #1\expandafter \@firstoftwo
 \else \expandafter \@secondoftwo
 \fi
}%
\providecommand \@ifx [1]{%
 \ifx #1\expandafter \@firstoftwo
 \else \expandafter \@secondoftwo
 \fi
}%
\providecommand \natexlab [1]{#1}%
\providecommand \enquote  [1]{``#1''}%
\providecommand \bibnamefont  [1]{#1}%
\providecommand \bibfnamefont [1]{#1}%
\providecommand \citenamefont [1]{#1}%
\providecommand \href@noop [0]{\@secondoftwo}%
\providecommand \href [0]{\begingroup \@sanitize@url \@href}%
\providecommand \@href[1]{\@@startlink{#1}\@@href}%
\providecommand \@@href[1]{\endgroup#1\@@endlink}%
\providecommand \@sanitize@url [0]{\catcode `\\12\catcode `\$12\catcode
  `\&12\catcode `\#12\catcode `\^12\catcode `\_12\catcode `\%12\relax}%
\providecommand \@@startlink[1]{}%
\providecommand \@@endlink[0]{}%
\providecommand \url  [0]{\begingroup\@sanitize@url \@url }%
\providecommand \@url [1]{\endgroup\@href {#1}{\urlprefix }}%
\providecommand \urlprefix  [0]{URL }%
\providecommand \Eprint [0]{\href }%
\providecommand \doibase [0]{https://doi.org/}%
\providecommand \selectlanguage [0]{\@gobble}%
\providecommand \bibinfo  [0]{\@secondoftwo}%
\providecommand \bibfield  [0]{\@secondoftwo}%
\providecommand \translation [1]{[#1]}%
\providecommand \BibitemOpen [0]{}%
\providecommand \bibitemStop [0]{}%
\providecommand \bibitemNoStop [0]{.\EOS\space}%
\providecommand \EOS [0]{\spacefactor3000\relax}%
\providecommand \BibitemShut  [1]{\csname bibitem#1\endcsname}%
\let\auto@bib@innerbib\@empty
\bibitem [{\citenamefont {Wendt}(2008)}]{wendt2008computational}%
  \BibitemOpen
  \bibfield  {author} {\bibinfo {author} {\bibfnamefont {J.~F.}\ \bibnamefont
  {Wendt}},\ }\href@noop {} {\emph {\bibinfo {title} {Computational fluid
  dynamics: an introduction}}}\ (\bibinfo  {publisher} {Springer Science \&
  Business Media},\ \bibinfo {year} {2008})\BibitemShut {NoStop}%
\bibitem [{\citenamefont {Slotnick}\ \emph {et~al.}(2014)\citenamefont
  {Slotnick}, \citenamefont {Khodadoust}, \citenamefont {Alonso}, \citenamefont
  {Darmofal}, \citenamefont {Gropp}, \citenamefont {Lurie},\ and\ \citenamefont
  {Mavriplis}}]{slotnick2014cfd}%
  \BibitemOpen
  \bibfield  {author} {\bibinfo {author} {\bibfnamefont {J.~P.}\ \bibnamefont
  {Slotnick}}, \bibinfo {author} {\bibfnamefont {A.}~\bibnamefont
  {Khodadoust}}, \bibinfo {author} {\bibfnamefont {J.}~\bibnamefont {Alonso}},
  \bibinfo {author} {\bibfnamefont {D.}~\bibnamefont {Darmofal}}, \bibinfo
  {author} {\bibfnamefont {W.}~\bibnamefont {Gropp}}, \bibinfo {author}
  {\bibfnamefont {E.}~\bibnamefont {Lurie}},\ and\ \bibinfo {author}
  {\bibfnamefont {D.~J.}\ \bibnamefont {Mavriplis}},\ }\href@noop {} {\emph
  {\bibinfo {title} {CFD vision 2030 study: a path to revolutionary
  computational aerosciences}}},\ \bibinfo {type} {Tech. Rep.}\ (\bibinfo
  {year} {2014})\BibitemShut {NoStop}%
\bibitem [{\citenamefont {Drikakis}\ \emph {et~al.}(2019)\citenamefont
  {Drikakis}, \citenamefont {Frank},\ and\ \citenamefont
  {Tabor}}]{drikakis2019multiscale}%
  \BibitemOpen
  \bibfield  {author} {\bibinfo {author} {\bibfnamefont {D.}~\bibnamefont
  {Drikakis}}, \bibinfo {author} {\bibfnamefont {M.}~\bibnamefont {Frank}},\
  and\ \bibinfo {author} {\bibfnamefont {G.}~\bibnamefont {Tabor}},\ }\bibfield
   {title} {\bibinfo {title} {Multiscale computational fluid dynamics},\
  }\href@noop {} {\bibfield  {journal} {\bibinfo  {journal} {Energies}\
  }\textbf {\bibinfo {volume} {12}},\ \bibinfo {pages} {3272} (\bibinfo {year}
  {2019})}\BibitemShut {NoStop}%
\bibitem [{\citenamefont {Trebotich}(2024)}]{trebotich2024exascale}%
  \BibitemOpen
  \bibfield  {author} {\bibinfo {author} {\bibfnamefont {D.}~\bibnamefont
  {Trebotich}},\ }\bibfield  {title} {\bibinfo {title} {Exascale computational
  fluid dynamics in heterogeneous systems},\ }\href@noop {} {\bibfield
  {journal} {\bibinfo  {journal} {Journal of Fluids Engineering}\ }\textbf
  {\bibinfo {volume} {146}} (\bibinfo {year} {2024})}\BibitemShut {NoStop}%
\bibitem [{\citenamefont {Bhatti}\ \emph {et~al.}(2020)\citenamefont {Bhatti},
  \citenamefont {Marin}, \citenamefont {Zeeshan},\ and\ \citenamefont
  {Abdelsalam}}]{bhatti2020recent}%
  \BibitemOpen
  \bibfield  {author} {\bibinfo {author} {\bibfnamefont {M.~M.}\ \bibnamefont
  {Bhatti}}, \bibinfo {author} {\bibfnamefont {M.}~\bibnamefont {Marin}},
  \bibinfo {author} {\bibfnamefont {A.}~\bibnamefont {Zeeshan}},\ and\ \bibinfo
  {author} {\bibfnamefont {S.~I.}\ \bibnamefont {Abdelsalam}},\ }\bibfield
  {title} {\bibinfo {title} {Recent trends in computational fluid dynamics},\
  }\href@noop {} {\bibfield  {journal} {\bibinfo  {journal} {Frontiers in
  Physics}\ }\textbf {\bibinfo {volume} {8}},\ \bibinfo {pages} {593111}
  (\bibinfo {year} {2020})}\BibitemShut {NoStop}%
\bibitem [{\citenamefont {Constantin}\ and\ \citenamefont
  {Foia{\c{s}}}(1988)}]{constantin1988navier}%
  \BibitemOpen
  \bibfield  {author} {\bibinfo {author} {\bibfnamefont {P.}~\bibnamefont
  {Constantin}}\ and\ \bibinfo {author} {\bibfnamefont {C.}~\bibnamefont
  {Foia{\c{s}}}},\ }\href@noop {} {\emph {\bibinfo {title} {Navier-stokes
  equations}}}\ (\bibinfo  {publisher} {University of Chicago press},\ \bibinfo
  {year} {1988})\BibitemShut {NoStop}%
\bibitem [{\citenamefont {Succi}\ \emph {et~al.}(2023)\citenamefont {Succi},
  \citenamefont {Itani}, \citenamefont {Sreenivasan},\ and\ \citenamefont
  {Steijl}}]{succi2023quantum}%
  \BibitemOpen
  \bibfield  {author} {\bibinfo {author} {\bibfnamefont {S.}~\bibnamefont
  {Succi}}, \bibinfo {author} {\bibfnamefont {W.}~\bibnamefont {Itani}},
  \bibinfo {author} {\bibfnamefont {K.}~\bibnamefont {Sreenivasan}},\ and\
  \bibinfo {author} {\bibfnamefont {R.}~\bibnamefont {Steijl}},\ }\bibfield
  {title} {\bibinfo {title} {Quantum computing for fluids: Where do we
  stand?},\ }\href@noop {} {\bibfield  {journal} {\bibinfo  {journal}
  {Europhysics Letters}\ }\textbf {\bibinfo {volume} {144}},\ \bibinfo {pages}
  {10001} (\bibinfo {year} {2023})}\BibitemShut {NoStop}%
\bibitem [{\citenamefont {Nejadmalayeri}\ \emph {et~al.}(2013)\citenamefont
  {Nejadmalayeri}, \citenamefont {Vezolainen},\ and\ \citenamefont
  {Vasilyev}}]{nejadmalayeri2013reynolds}%
  \BibitemOpen
  \bibfield  {author} {\bibinfo {author} {\bibfnamefont {A.}~\bibnamefont
  {Nejadmalayeri}}, \bibinfo {author} {\bibfnamefont {A.}~\bibnamefont
  {Vezolainen}},\ and\ \bibinfo {author} {\bibfnamefont {O.~V.}\ \bibnamefont
  {Vasilyev}},\ }\bibfield  {title} {\bibinfo {title} {Reynolds number scaling
  of coherent vortex simulation and stochastic coherent adaptive large eddy
  simulation},\ }\href@noop {} {\bibfield  {journal} {\bibinfo  {journal}
  {Physics of Fluids}\ }\textbf {\bibinfo {volume} {25}} (\bibinfo {year}
  {2013})}\BibitemShut {NoStop}%
\bibitem [{\citenamefont {Harrow}\ \emph {et~al.}(2009)\citenamefont {Harrow},
  \citenamefont {Hassidim},\ and\ \citenamefont {Lloyd}}]{harrow2009quantum}%
  \BibitemOpen
  \bibfield  {author} {\bibinfo {author} {\bibfnamefont {A.~W.}\ \bibnamefont
  {Harrow}}, \bibinfo {author} {\bibfnamefont {A.}~\bibnamefont {Hassidim}},\
  and\ \bibinfo {author} {\bibfnamefont {S.}~\bibnamefont {Lloyd}},\ }\bibfield
   {title} {\bibinfo {title} {Quantum algorithm for linear systems of
  equations},\ }\href@noop {} {\bibfield  {journal} {\bibinfo  {journal}
  {Physical review letters}\ }\textbf {\bibinfo {volume} {103}},\ \bibinfo
  {pages} {150502} (\bibinfo {year} {2009})}\BibitemShut {NoStop}%
\bibitem [{\citenamefont {Low}\ and\ \citenamefont
  {Chuang}(2017)}]{low2017optimal}%
  \BibitemOpen
  \bibfield  {author} {\bibinfo {author} {\bibfnamefont {G.~H.}\ \bibnamefont
  {Low}}\ and\ \bibinfo {author} {\bibfnamefont {I.~L.}\ \bibnamefont
  {Chuang}},\ }\bibfield  {title} {\bibinfo {title} {Optimal hamiltonian
  simulation by quantum signal processing},\ }\href@noop {} {\bibfield
  {journal} {\bibinfo  {journal} {Physical review letters}\ }\textbf {\bibinfo
  {volume} {118}},\ \bibinfo {pages} {010501} (\bibinfo {year}
  {2017})}\BibitemShut {NoStop}%
\bibitem [{\citenamefont {Gily{\'e}n}\ \emph {et~al.}(2019)\citenamefont
  {Gily{\'e}n}, \citenamefont {Su}, \citenamefont {Low},\ and\ \citenamefont
  {Wiebe}}]{gilyen2019quantum}%
  \BibitemOpen
  \bibfield  {author} {\bibinfo {author} {\bibfnamefont {A.}~\bibnamefont
  {Gily{\'e}n}}, \bibinfo {author} {\bibfnamefont {Y.}~\bibnamefont {Su}},
  \bibinfo {author} {\bibfnamefont {G.~H.}\ \bibnamefont {Low}},\ and\ \bibinfo
  {author} {\bibfnamefont {N.}~\bibnamefont {Wiebe}},\ }\bibfield  {title}
  {\bibinfo {title} {Quantum singular value transformation and beyond:
  exponential improvements for quantum matrix arithmetics},\ }in\ \href@noop {}
  {\emph {\bibinfo {booktitle} {Proceedings of the 51st Annual ACM SIGACT
  Symposium on Theory of Computing}}}\ (\bibinfo {year} {2019})\ pp.\ \bibinfo
  {pages} {193--204}\BibitemShut {NoStop}%
\bibitem [{\citenamefont {Berry}(2014)}]{berry2014high}%
  \BibitemOpen
  \bibfield  {author} {\bibinfo {author} {\bibfnamefont {D.~W.}\ \bibnamefont
  {Berry}},\ }\bibfield  {title} {\bibinfo {title} {High-order quantum
  algorithm for solving linear differential equations},\ }\href@noop {}
  {\bibfield  {journal} {\bibinfo  {journal} {Journal of Physics A:
  Mathematical and Theoretical}\ }\textbf {\bibinfo {volume} {47}},\ \bibinfo
  {pages} {105301} (\bibinfo {year} {2014})}\BibitemShut {NoStop}%
\bibitem [{\citenamefont {Berry}\ \emph {et~al.}(2017)\citenamefont {Berry},
  \citenamefont {Childs}, \citenamefont {Ostrander},\ and\ \citenamefont
  {Wang}}]{berry2017quantum}%
  \BibitemOpen
  \bibfield  {author} {\bibinfo {author} {\bibfnamefont {D.~W.}\ \bibnamefont
  {Berry}}, \bibinfo {author} {\bibfnamefont {A.~M.}\ \bibnamefont {Childs}},
  \bibinfo {author} {\bibfnamefont {A.}~\bibnamefont {Ostrander}},\ and\
  \bibinfo {author} {\bibfnamefont {G.}~\bibnamefont {Wang}},\ }\bibfield
  {title} {\bibinfo {title} {Quantum algorithm for linear differential
  equations with exponentially improved dependence on precision},\ }\href@noop
  {} {\bibfield  {journal} {\bibinfo  {journal} {Communications in Mathematical
  Physics}\ }\textbf {\bibinfo {volume} {356}},\ \bibinfo {pages} {1057}
  (\bibinfo {year} {2017})}\BibitemShut {NoStop}%
\bibitem [{\citenamefont {Xin}\ \emph {et~al.}(2020)\citenamefont {Xin},
  \citenamefont {Wei}, \citenamefont {Cui}, \citenamefont {Xiao}, \citenamefont
  {Arrazola}, \citenamefont {Lamata}, \citenamefont {Kong}, \citenamefont {Lu},
  \citenamefont {Solano},\ and\ \citenamefont {Long}}]{xin2020quantum}%
  \BibitemOpen
  \bibfield  {author} {\bibinfo {author} {\bibfnamefont {T.}~\bibnamefont
  {Xin}}, \bibinfo {author} {\bibfnamefont {S.}~\bibnamefont {Wei}}, \bibinfo
  {author} {\bibfnamefont {J.}~\bibnamefont {Cui}}, \bibinfo {author}
  {\bibfnamefont {J.}~\bibnamefont {Xiao}}, \bibinfo {author} {\bibfnamefont
  {I.}~\bibnamefont {Arrazola}}, \bibinfo {author} {\bibfnamefont
  {L.}~\bibnamefont {Lamata}}, \bibinfo {author} {\bibfnamefont
  {X.}~\bibnamefont {Kong}}, \bibinfo {author} {\bibfnamefont {D.}~\bibnamefont
  {Lu}}, \bibinfo {author} {\bibfnamefont {E.}~\bibnamefont {Solano}},\ and\
  \bibinfo {author} {\bibfnamefont {G.}~\bibnamefont {Long}},\ }\bibfield
  {title} {\bibinfo {title} {Quantum algorithm for solving linear differential
  equations: Theory and experiment},\ }\href@noop {} {\bibfield  {journal}
  {\bibinfo  {journal} {Physical Review A}\ }\textbf {\bibinfo {volume}
  {101}},\ \bibinfo {pages} {032307} (\bibinfo {year} {2020})}\BibitemShut
  {NoStop}%
\bibitem [{\citenamefont {Liu}\ \emph {et~al.}(2021)\citenamefont {Liu},
  \citenamefont {Kolden}, \citenamefont {Krovi}, \citenamefont {Loureiro},
  \citenamefont {Trivisa},\ and\ \citenamefont {Childs}}]{liu2021efficient}%
  \BibitemOpen
  \bibfield  {author} {\bibinfo {author} {\bibfnamefont {J.-P.}\ \bibnamefont
  {Liu}}, \bibinfo {author} {\bibfnamefont {H.~{\O}.}\ \bibnamefont {Kolden}},
  \bibinfo {author} {\bibfnamefont {H.~K.}\ \bibnamefont {Krovi}}, \bibinfo
  {author} {\bibfnamefont {N.~F.}\ \bibnamefont {Loureiro}}, \bibinfo {author}
  {\bibfnamefont {K.}~\bibnamefont {Trivisa}},\ and\ \bibinfo {author}
  {\bibfnamefont {A.~M.}\ \bibnamefont {Childs}},\ }\bibfield  {title}
  {\bibinfo {title} {Efficient quantum algorithm for dissipative nonlinear
  differential equations},\ }\href@noop {} {\bibfield  {journal} {\bibinfo
  {journal} {Proceedings of the National Academy of Sciences}\ }\textbf
  {\bibinfo {volume} {118}},\ \bibinfo {pages} {e2026805118} (\bibinfo {year}
  {2021})}\BibitemShut {NoStop}%
\bibitem [{\citenamefont {Wang}\ \emph {et~al.}(2020)\citenamefont {Wang},
  \citenamefont {Wang}, \citenamefont {Li}, \citenamefont {Fan}, \citenamefont
  {Wei},\ and\ \citenamefont {Gu}}]{wang2020quantum}%
  \BibitemOpen
  \bibfield  {author} {\bibinfo {author} {\bibfnamefont {S.}~\bibnamefont
  {Wang}}, \bibinfo {author} {\bibfnamefont {Z.}~\bibnamefont {Wang}}, \bibinfo
  {author} {\bibfnamefont {W.}~\bibnamefont {Li}}, \bibinfo {author}
  {\bibfnamefont {L.}~\bibnamefont {Fan}}, \bibinfo {author} {\bibfnamefont
  {Z.}~\bibnamefont {Wei}},\ and\ \bibinfo {author} {\bibfnamefont
  {Y.}~\bibnamefont {Gu}},\ }\bibfield  {title} {\bibinfo {title} {Quantum fast
  poisson solver: the algorithm and complete and modular circuit design},\
  }\href@noop {} {\bibfield  {journal} {\bibinfo  {journal} {Quantum
  Information Processing}\ }\textbf {\bibinfo {volume} {19}},\ \bibinfo {pages}
  {1} (\bibinfo {year} {2020})}\BibitemShut {NoStop}%
\bibitem [{\citenamefont {Chen}\ \emph {et~al.}(2024)\citenamefont {Chen},
  \citenamefont {Ma}, \citenamefont {Ye}, \citenamefont {Xu}, \citenamefont
  {Bai}, \citenamefont {Zhou}, \citenamefont {Tan}, \citenamefont {Zhuang},
  \citenamefont {Xu}, \citenamefont {Wang} \emph {et~al.}}]{chen2024enabling}%
  \BibitemOpen
  \bibfield  {author} {\bibinfo {author} {\bibfnamefont {Z.-Y.}\ \bibnamefont
  {Chen}}, \bibinfo {author} {\bibfnamefont {T.-Y.}\ \bibnamefont {Ma}},
  \bibinfo {author} {\bibfnamefont {C.-C.}\ \bibnamefont {Ye}}, \bibinfo
  {author} {\bibfnamefont {L.}~\bibnamefont {Xu}}, \bibinfo {author}
  {\bibfnamefont {W.}~\bibnamefont {Bai}}, \bibinfo {author} {\bibfnamefont
  {L.}~\bibnamefont {Zhou}}, \bibinfo {author} {\bibfnamefont {M.-Y.}\
  \bibnamefont {Tan}}, \bibinfo {author} {\bibfnamefont {X.-N.}\ \bibnamefont
  {Zhuang}}, \bibinfo {author} {\bibfnamefont {X.-F.}\ \bibnamefont {Xu}},
  \bibinfo {author} {\bibfnamefont {Y.-J.}\ \bibnamefont {Wang}}, \emph
  {et~al.},\ }\bibfield  {title} {\bibinfo {title} {Enabling large-scale and
  high-precision fluid simulations on near-term quantum computers},\
  }\href@noop {} {\bibfield  {journal} {\bibinfo  {journal} {Computer Methods
  in Applied Mechanics and Engineering}\ }\textbf {\bibinfo {volume} {432}},\
  \bibinfo {pages} {117428} (\bibinfo {year} {2024})}\BibitemShut {NoStop}%
\bibitem [{\citenamefont {Kupershmidt}\ and\ \citenamefont
  {Mathieu}(1989)}]{kupershmidt1989quantum}%
  \BibitemOpen
  \bibfield  {author} {\bibinfo {author} {\bibfnamefont {B.}~\bibnamefont
  {Kupershmidt}}\ and\ \bibinfo {author} {\bibfnamefont {P.}~\bibnamefont
  {Mathieu}},\ }\bibfield  {title} {\bibinfo {title} {Quantum korteweg-de vries
  like equations and perturbed conformal field theories},\ }\href@noop {}
  {\bibfield  {journal} {\bibinfo  {journal} {Physics Letters B}\ }\textbf
  {\bibinfo {volume} {227}},\ \bibinfo {pages} {245} (\bibinfo {year}
  {1989})}\BibitemShut {NoStop}%
\bibitem [{\citenamefont {Aaronson}(2015)}]{aaronson2015read}%
  \BibitemOpen
  \bibfield  {author} {\bibinfo {author} {\bibfnamefont {S.}~\bibnamefont
  {Aaronson}},\ }\bibfield  {title} {\bibinfo {title} {Read the fine print},\
  }\href@noop {} {\bibfield  {journal} {\bibinfo  {journal} {Nature Physics}\
  }\textbf {\bibinfo {volume} {11}},\ \bibinfo {pages} {291} (\bibinfo {year}
  {2015})}\BibitemShut {NoStop}%
\bibitem [{\citenamefont {Berkooz}\ \emph {et~al.}(1993)\citenamefont
  {Berkooz}, \citenamefont {Holmes},\ and\ \citenamefont
  {Lumley}}]{berkooz1993proper}%
  \BibitemOpen
  \bibfield  {author} {\bibinfo {author} {\bibfnamefont {G.}~\bibnamefont
  {Berkooz}}, \bibinfo {author} {\bibfnamefont {P.}~\bibnamefont {Holmes}},\
  and\ \bibinfo {author} {\bibfnamefont {J.~L.}\ \bibnamefont {Lumley}},\
  }\bibfield  {title} {\bibinfo {title} {The proper orthogonal decomposition in
  the analysis of turbulent flows},\ }\href@noop {} {\bibfield  {journal}
  {\bibinfo  {journal} {Annual review of fluid mechanics}\ }\textbf {\bibinfo
  {volume} {25}},\ \bibinfo {pages} {539} (\bibinfo {year} {1993})}\BibitemShut
  {NoStop}%
\bibitem [{\citenamefont {Schmid}(2010)}]{schmid2010dynamic}%
  \BibitemOpen
  \bibfield  {author} {\bibinfo {author} {\bibfnamefont {P.~J.}\ \bibnamefont
  {Schmid}},\ }\bibfield  {title} {\bibinfo {title} {Dynamic mode decomposition
  of numerical and experimental data},\ }\href@noop {} {\bibfield  {journal}
  {\bibinfo  {journal} {Journal of fluid mechanics}\ }\textbf {\bibinfo
  {volume} {656}},\ \bibinfo {pages} {5} (\bibinfo {year} {2010})}\BibitemShut
  {NoStop}%
\bibitem [{\citenamefont {Lloyd}\ \emph {et~al.}(2014)\citenamefont {Lloyd},
  \citenamefont {Mohseni},\ and\ \citenamefont {Rebentrost}}]{QPCA2014}%
  \BibitemOpen
  \bibfield  {author} {\bibinfo {author} {\bibfnamefont {S.}~\bibnamefont
  {Lloyd}}, \bibinfo {author} {\bibfnamefont {M.}~\bibnamefont {Mohseni}},\
  and\ \bibinfo {author} {\bibfnamefont {P.}~\bibnamefont {Rebentrost}},\
  }\bibfield  {title} {\bibinfo {title} {Quantum principal component
  analysis},\ }\href@noop {} {\bibfield  {journal} {\bibinfo  {journal} {Nature
  Physics}\ }\textbf {\bibinfo {volume} {10}},\ \bibinfo {pages} {631}
  (\bibinfo {year} {2014})}\BibitemShut {NoStop}%
\bibitem [{\citenamefont {Martyn}\ \emph {et~al.}(2021)\citenamefont {Martyn},
  \citenamefont {Rossi}, \citenamefont {Tan},\ and\ \citenamefont
  {Chuang}}]{martyn2021grand}%
  \BibitemOpen
  \bibfield  {author} {\bibinfo {author} {\bibfnamefont {J.~M.}\ \bibnamefont
  {Martyn}}, \bibinfo {author} {\bibfnamefont {Z.~M.}\ \bibnamefont {Rossi}},
  \bibinfo {author} {\bibfnamefont {A.~K.}\ \bibnamefont {Tan}},\ and\ \bibinfo
  {author} {\bibfnamefont {I.~L.}\ \bibnamefont {Chuang}},\ }\bibfield  {title}
  {\bibinfo {title} {Grand unification of quantum algorithms},\ }\href@noop {}
  {\bibfield  {journal} {\bibinfo  {journal} {PRX quantum}\ }\textbf {\bibinfo
  {volume} {2}},\ \bibinfo {pages} {040203} (\bibinfo {year}
  {2021})}\BibitemShut {NoStop}%
\bibitem [{\citenamefont {Cerezo}\ \emph {et~al.}(2021)\citenamefont {Cerezo},
  \citenamefont {Arrasmith}, \citenamefont {Babbush}, \citenamefont {Benjamin},
  \citenamefont {Endo}, \citenamefont {Fujii}, \citenamefont {McClean},
  \citenamefont {Mitarai}, \citenamefont {Yuan}, \citenamefont {Cincio} \emph
  {et~al.}}]{cerezo2021variational}%
  \BibitemOpen
  \bibfield  {author} {\bibinfo {author} {\bibfnamefont {M.}~\bibnamefont
  {Cerezo}}, \bibinfo {author} {\bibfnamefont {A.}~\bibnamefont {Arrasmith}},
  \bibinfo {author} {\bibfnamefont {R.}~\bibnamefont {Babbush}}, \bibinfo
  {author} {\bibfnamefont {S.~C.}\ \bibnamefont {Benjamin}}, \bibinfo {author}
  {\bibfnamefont {S.}~\bibnamefont {Endo}}, \bibinfo {author} {\bibfnamefont
  {K.}~\bibnamefont {Fujii}}, \bibinfo {author} {\bibfnamefont {J.~R.}\
  \bibnamefont {McClean}}, \bibinfo {author} {\bibfnamefont {K.}~\bibnamefont
  {Mitarai}}, \bibinfo {author} {\bibfnamefont {X.}~\bibnamefont {Yuan}},
  \bibinfo {author} {\bibfnamefont {L.}~\bibnamefont {Cincio}}, \emph
  {et~al.},\ }\bibfield  {title} {\bibinfo {title} {Variational quantum
  algorithms},\ }\href@noop {} {\bibfield  {journal} {\bibinfo  {journal}
  {Nature Reviews Physics}\ }\textbf {\bibinfo {volume} {3}},\ \bibinfo {pages}
  {625} (\bibinfo {year} {2021})}\BibitemShut {NoStop}%
\bibitem [{\citenamefont {Williams}\ \emph {et~al.}(2024)\citenamefont
  {Williams}, \citenamefont {Scali}, \citenamefont {Gentile}, \citenamefont
  {Berger},\ and\ \citenamefont {Kyriienko}}]{williams2024addressing}%
  \BibitemOpen
  \bibfield  {author} {\bibinfo {author} {\bibfnamefont {C.~A.}\ \bibnamefont
  {Williams}}, \bibinfo {author} {\bibfnamefont {S.}~\bibnamefont {Scali}},
  \bibinfo {author} {\bibfnamefont {A.~A.}\ \bibnamefont {Gentile}}, \bibinfo
  {author} {\bibfnamefont {D.}~\bibnamefont {Berger}},\ and\ \bibinfo {author}
  {\bibfnamefont {O.}~\bibnamefont {Kyriienko}},\ }\bibfield  {title} {\bibinfo
  {title} {Addressing the readout problem in quantum differential equation
  algorithms with quantum scientific machine learning},\ }\href@noop {}
  {\bibfield  {journal} {\bibinfo  {journal} {arXiv preprint arXiv:2411.14259}\
  } (\bibinfo {year} {2024})}\BibitemShut {NoStop}%
\bibitem [{\citenamefont {Rebentrost}\ \emph {et~al.}(2014)\citenamefont
  {Rebentrost}, \citenamefont {Mohseni},\ and\ \citenamefont
  {Lloyd}}]{rebentrost2014quantum}%
  \BibitemOpen
  \bibfield  {author} {\bibinfo {author} {\bibfnamefont {P.}~\bibnamefont
  {Rebentrost}}, \bibinfo {author} {\bibfnamefont {M.}~\bibnamefont
  {Mohseni}},\ and\ \bibinfo {author} {\bibfnamefont {S.}~\bibnamefont
  {Lloyd}},\ }\bibfield  {title} {\bibinfo {title} {Quantum support vector
  machine for big data classification},\ }\href@noop {} {\bibfield  {journal}
  {\bibinfo  {journal} {Physical review letters}\ }\textbf {\bibinfo {volume}
  {113}},\ \bibinfo {pages} {130503} (\bibinfo {year} {2014})}\BibitemShut
  {NoStop}%
\bibitem [{\citenamefont {Yuan}\ \emph {et~al.}(2023)\citenamefont {Yuan},
  \citenamefont {Chen}, \citenamefont {Liu}, \citenamefont {Xie}, \citenamefont
  {Liu}, \citenamefont {Jin}, \citenamefont {Wen},\ and\ \citenamefont
  {Tang}}]{yuan2023quantum}%
  \BibitemOpen
  \bibfield  {author} {\bibinfo {author} {\bibfnamefont {X.-J.}\ \bibnamefont
  {Yuan}}, \bibinfo {author} {\bibfnamefont {Z.-Q.}\ \bibnamefont {Chen}},
  \bibinfo {author} {\bibfnamefont {Y.-D.}\ \bibnamefont {Liu}}, \bibinfo
  {author} {\bibfnamefont {Z.}~\bibnamefont {Xie}}, \bibinfo {author}
  {\bibfnamefont {Y.-Z.}\ \bibnamefont {Liu}}, \bibinfo {author} {\bibfnamefont
  {X.-M.}\ \bibnamefont {Jin}}, \bibinfo {author} {\bibfnamefont
  {X.}~\bibnamefont {Wen}},\ and\ \bibinfo {author} {\bibfnamefont
  {H.}~\bibnamefont {Tang}},\ }\bibfield  {title} {\bibinfo {title} {Quantum
  support vector machines for aerodynamic classification},\ }\href@noop {}
  {\bibfield  {journal} {\bibinfo  {journal} {Intelligent Computing}\ }\textbf
  {\bibinfo {volume} {2}},\ \bibinfo {pages} {0057} (\bibinfo {year}
  {2023})}\BibitemShut {NoStop}%
\bibitem [{\citenamefont {Nivelkar}\ \emph {et~al.}(2023)\citenamefont
  {Nivelkar}, \citenamefont {Bhirud}, \citenamefont {Singh}, \citenamefont
  {Ranjan},\ and\ \citenamefont {Kumar}}]{nivelkar2023quantum}%
  \BibitemOpen
  \bibfield  {author} {\bibinfo {author} {\bibfnamefont {M.}~\bibnamefont
  {Nivelkar}}, \bibinfo {author} {\bibfnamefont {S.}~\bibnamefont {Bhirud}},
  \bibinfo {author} {\bibfnamefont {M.}~\bibnamefont {Singh}}, \bibinfo
  {author} {\bibfnamefont {R.}~\bibnamefont {Ranjan}},\ and\ \bibinfo {author}
  {\bibfnamefont {B.}~\bibnamefont {Kumar}},\ }\bibfield  {title} {\bibinfo
  {title} {Quantum computing to study cloud turbulence properties},\
  }\href@noop {} {\bibfield  {journal} {\bibinfo  {journal} {IEEE Access}\
  }\textbf {\bibinfo {volume} {11}},\ \bibinfo {pages} {70679} (\bibinfo {year}
  {2023})}\BibitemShut {NoStop}%
\bibitem [{\citenamefont {Xue}\ \emph {et~al.}(2023)\citenamefont {Xue},
  \citenamefont {Chen}, \citenamefont {Sun}, \citenamefont {Xu}, \citenamefont
  {Chen}, \citenamefont {Liu}, \citenamefont {Zhuang}, \citenamefont {Wu},\
  and\ \citenamefont {Guo}}]{xue2023quantum}%
  \BibitemOpen
  \bibfield  {author} {\bibinfo {author} {\bibfnamefont {C.}~\bibnamefont
  {Xue}}, \bibinfo {author} {\bibfnamefont {Z.-Y.}\ \bibnamefont {Chen}},
  \bibinfo {author} {\bibfnamefont {T.-P.}\ \bibnamefont {Sun}}, \bibinfo
  {author} {\bibfnamefont {X.-F.}\ \bibnamefont {Xu}}, \bibinfo {author}
  {\bibfnamefont {S.-M.}\ \bibnamefont {Chen}}, \bibinfo {author}
  {\bibfnamefont {H.-Y.}\ \bibnamefont {Liu}}, \bibinfo {author} {\bibfnamefont
  {X.-N.}\ \bibnamefont {Zhuang}}, \bibinfo {author} {\bibfnamefont {Y.-C.}\
  \bibnamefont {Wu}},\ and\ \bibinfo {author} {\bibfnamefont {G.-P.}\
  \bibnamefont {Guo}},\ }\bibfield  {title} {\bibinfo {title} {Quantum dynamic
  mode decomposition algorithm for high-dimensional time series analysis},\
  }\href@noop {} {\bibfield  {journal} {\bibinfo  {journal} {Intelligent
  Computing}\ }\textbf {\bibinfo {volume} {2}},\ \bibinfo {pages} {0045}
  (\bibinfo {year} {2023})}\BibitemShut {NoStop}%
\bibitem [{\citenamefont {Jaksch}\ \emph {et~al.}(2023)\citenamefont {Jaksch},
  \citenamefont {Givi}, \citenamefont {Daley},\ and\ \citenamefont
  {Rung}}]{jaksch2023variational}%
  \BibitemOpen
  \bibfield  {author} {\bibinfo {author} {\bibfnamefont {D.}~\bibnamefont
  {Jaksch}}, \bibinfo {author} {\bibfnamefont {P.}~\bibnamefont {Givi}},
  \bibinfo {author} {\bibfnamefont {A.~J.}\ \bibnamefont {Daley}},\ and\
  \bibinfo {author} {\bibfnamefont {T.}~\bibnamefont {Rung}},\ }\bibfield
  {title} {\bibinfo {title} {Variational quantum algorithms for computational
  fluid dynamics},\ }\href@noop {} {\bibfield  {journal} {\bibinfo  {journal}
  {AIAA journal}\ }\textbf {\bibinfo {volume} {61}},\ \bibinfo {pages} {1885}
  (\bibinfo {year} {2023})}\BibitemShut {NoStop}%
\bibitem [{\citenamefont {Yadav}(2023)}]{yadav2023qpde}%
  \BibitemOpen
  \bibfield  {author} {\bibinfo {author} {\bibfnamefont {S.}~\bibnamefont
  {Yadav}},\ }\bibfield  {title} {\bibinfo {title} {Qpde: quantum neural
  network based stabilization parameter prediction for numerical solvers for
  partial differential equations},\ }\href@noop {} {\bibfield  {journal}
  {\bibinfo  {journal} {AppliedMath}\ }\textbf {\bibinfo {volume} {3}},\
  \bibinfo {pages} {552} (\bibinfo {year} {2023})}\BibitemShut {NoStop}%
\bibitem [{\citenamefont {Bazgir}(2024)}]{bazgir2024hybrid}%
  \BibitemOpen
  \bibfield  {author} {\bibinfo {author} {\bibfnamefont {A.}~\bibnamefont
  {Bazgir}},\ }\emph {\bibinfo {title} {Hybrid Quantum-Classical Machine
  Learning for Canonical Fluid Dynamics and Heat Transfer Problems}},\
  \href@noop {} {Master's thesis},\ \bibinfo  {school} {University of
  Missouri-Columbia} (\bibinfo {year} {2024})\BibitemShut {NoStop}%
\bibitem [{\citenamefont {Sedykh}\ \emph {et~al.}(2024)\citenamefont {Sedykh},
  \citenamefont {Podapaka}, \citenamefont {Sagingalieva}, \citenamefont
  {Pinto}, \citenamefont {Pflitsch},\ and\ \citenamefont
  {Melnikov}}]{sedykh2024hybrid}%
  \BibitemOpen
  \bibfield  {author} {\bibinfo {author} {\bibfnamefont {A.}~\bibnamefont
  {Sedykh}}, \bibinfo {author} {\bibfnamefont {M.}~\bibnamefont {Podapaka}},
  \bibinfo {author} {\bibfnamefont {A.}~\bibnamefont {Sagingalieva}}, \bibinfo
  {author} {\bibfnamefont {K.}~\bibnamefont {Pinto}}, \bibinfo {author}
  {\bibfnamefont {M.}~\bibnamefont {Pflitsch}},\ and\ \bibinfo {author}
  {\bibfnamefont {A.}~\bibnamefont {Melnikov}},\ }\bibfield  {title} {\bibinfo
  {title} {Hybrid quantum physics-informed neural networks for simulating
  computational fluid dynamics in complex shapes},\ }\href@noop {} {\bibfield
  {journal} {\bibinfo  {journal} {Machine Learning: Science and Technology}\
  }\textbf {\bibinfo {volume} {5}},\ \bibinfo {pages} {025045} (\bibinfo {year}
  {2024})}\BibitemShut {NoStop}%
\bibitem [{\citenamefont {Liu}\ \emph {et~al.}(2023)\citenamefont {Liu},
  \citenamefont {Sun}, \citenamefont {Wu}, \citenamefont {Han},\ and\
  \citenamefont {Guo}}]{liu2023mitigating}%
  \BibitemOpen
  \bibfield  {author} {\bibinfo {author} {\bibfnamefont {H.-Y.}\ \bibnamefont
  {Liu}}, \bibinfo {author} {\bibfnamefont {T.-P.}\ \bibnamefont {Sun}},
  \bibinfo {author} {\bibfnamefont {Y.-C.}\ \bibnamefont {Wu}}, \bibinfo
  {author} {\bibfnamefont {Y.-J.}\ \bibnamefont {Han}},\ and\ \bibinfo {author}
  {\bibfnamefont {G.-P.}\ \bibnamefont {Guo}},\ }\bibfield  {title} {\bibinfo
  {title} {Mitigating barren plateaus with transfer-learning-inspired parameter
  initializations},\ }\href@noop {} {\bibfield  {journal} {\bibinfo  {journal}
  {New Journal of Physics}\ }\textbf {\bibinfo {volume} {25}},\ \bibinfo
  {pages} {013039} (\bibinfo {year} {2023})}\BibitemShut {NoStop}%
\bibitem [{\citenamefont {Szeg}(1939)}]{szeg1939orthogonal}%
  \BibitemOpen
  \bibfield  {author} {\bibinfo {author} {\bibfnamefont {G.}~\bibnamefont
  {Szeg}},\ }\href@noop {} {\emph {\bibinfo {title} {Orthogonal
  polynomials}}},\ Vol.~\bibinfo {volume} {23}\ (\bibinfo  {publisher}
  {American Mathematical Soc.},\ \bibinfo {year} {1939})\BibitemShut {NoStop}%
\bibitem [{\citenamefont {Nielsen}\ and\ \citenamefont
  {Chuang}(2010)}]{nielsen2010quantum}%
  \BibitemOpen
  \bibfield  {author} {\bibinfo {author} {\bibfnamefont {M.~A.}\ \bibnamefont
  {Nielsen}}\ and\ \bibinfo {author} {\bibfnamefont {I.~L.}\ \bibnamefont
  {Chuang}},\ }\href@noop {} {\emph {\bibinfo {title} {Quantum computation and
  quantum information}}}\ (\bibinfo  {publisher} {Cambridge university press},\
  \bibinfo {year} {2010})\BibitemShut {NoStop}%
\bibitem [{\citenamefont {Klappenecker}\ and\ \citenamefont
  {Rotteler}(2001)}]{klappenecker2001discrete}%
  \BibitemOpen
  \bibfield  {author} {\bibinfo {author} {\bibfnamefont {A.}~\bibnamefont
  {Klappenecker}}\ and\ \bibinfo {author} {\bibfnamefont {M.}~\bibnamefont
  {Rotteler}},\ }\bibfield  {title} {\bibinfo {title} {Discrete cosine
  transforms on quantum computers},\ }in\ \href@noop {} {\emph {\bibinfo
  {booktitle} {ISPA 2001. Proceedings of the 2nd International Symposium on
  Image and Signal Processing and Analysis. In conjunction with 23rd
  International Conference on Information Technology Interfaces (IEEE Cat.}}}\
  (\bibinfo {organization} {IEEE},\ \bibinfo {year} {2001})\ pp.\ \bibinfo
  {pages} {464--468}\BibitemShut {NoStop}%
\bibitem [{\citenamefont {Golub}\ \emph {et~al.}(1999)\citenamefont {Golub},
  \citenamefont {Hansen},\ and\ \citenamefont {O'Leary}}]{golub1999tikhonov}%
  \BibitemOpen
  \bibfield  {author} {\bibinfo {author} {\bibfnamefont {G.~H.}\ \bibnamefont
  {Golub}}, \bibinfo {author} {\bibfnamefont {P.~C.}\ \bibnamefont {Hansen}},\
  and\ \bibinfo {author} {\bibfnamefont {D.~P.}\ \bibnamefont {O'Leary}},\
  }\bibfield  {title} {\bibinfo {title} {Tikhonov regularization and total
  least squares},\ }\href@noop {} {\bibfield  {journal} {\bibinfo  {journal}
  {SIAM journal on matrix analysis and applications}\ }\textbf {\bibinfo
  {volume} {21}},\ \bibinfo {pages} {185} (\bibinfo {year} {1999})}\BibitemShut
  {NoStop}%
\bibitem [{\citenamefont {Luo}\ \emph {et~al.}(2023)\citenamefont {Luo},
  \citenamefont {Chen},\ and\ \citenamefont {Zhang}}]{luo2023cfdbench}%
  \BibitemOpen
  \bibfield  {author} {\bibinfo {author} {\bibfnamefont {Y.}~\bibnamefont
  {Luo}}, \bibinfo {author} {\bibfnamefont {Y.}~\bibnamefont {Chen}},\ and\
  \bibinfo {author} {\bibfnamefont {Z.}~\bibnamefont {Zhang}},\ }\bibfield
  {title} {\bibinfo {title} {Cfdbench: A large-scale benchmark for machine
  learning methods in fluid dynamics},\ }\href@noop {} {\bibfield  {journal}
  {\bibinfo  {journal} {arXiv preprint arXiv:2310.05963}\ } (\bibinfo {year}
  {2023})}\BibitemShut {NoStop}%
\bibitem [{\citenamefont {Dou}\ \emph {et~al.}(2022)\citenamefont {Dou},
  \citenamefont {Zou}, \citenamefont {Fang}, \citenamefont {Wang},
  \citenamefont {Zhao}, \citenamefont {Yu}, \citenamefont {Chen}, \citenamefont
  {Guo}, \citenamefont {Li}, \citenamefont {Chen} \emph
  {et~al.}}]{dou2022qpanda}%
  \BibitemOpen
  \bibfield  {author} {\bibinfo {author} {\bibfnamefont {M.}~\bibnamefont
  {Dou}}, \bibinfo {author} {\bibfnamefont {T.}~\bibnamefont {Zou}}, \bibinfo
  {author} {\bibfnamefont {Y.}~\bibnamefont {Fang}}, \bibinfo {author}
  {\bibfnamefont {J.}~\bibnamefont {Wang}}, \bibinfo {author} {\bibfnamefont
  {D.}~\bibnamefont {Zhao}}, \bibinfo {author} {\bibfnamefont {L.}~\bibnamefont
  {Yu}}, \bibinfo {author} {\bibfnamefont {B.}~\bibnamefont {Chen}}, \bibinfo
  {author} {\bibfnamefont {W.}~\bibnamefont {Guo}}, \bibinfo {author}
  {\bibfnamefont {Y.}~\bibnamefont {Li}}, \bibinfo {author} {\bibfnamefont
  {Z.}~\bibnamefont {Chen}}, \emph {et~al.},\ }\bibfield  {title} {\bibinfo
  {title} {Qpanda: high-performance quantum computing framework for multiple
  application scenarios},\ }\href@noop {} {\bibfield  {journal} {\bibinfo
  {journal} {arXiv preprint arXiv:2212.14201}\ } (\bibinfo {year}
  {2022})}\BibitemShut {NoStop}%
\bibitem [{\citenamefont {Bergholm}\ \emph {et~al.}(2018)\citenamefont
  {Bergholm}, \citenamefont {Izaac}, \citenamefont {Schuld}, \citenamefont
  {Gogolin}, \citenamefont {Ahmed}, \citenamefont {Ajith}, \citenamefont
  {Alam}, \citenamefont {Alonso-Linaje}, \citenamefont {AkashNarayanan},
  \citenamefont {Asadi} \emph {et~al.}}]{bergholm2018pennylane}%
  \BibitemOpen
  \bibfield  {author} {\bibinfo {author} {\bibfnamefont {V.}~\bibnamefont
  {Bergholm}}, \bibinfo {author} {\bibfnamefont {J.}~\bibnamefont {Izaac}},
  \bibinfo {author} {\bibfnamefont {M.}~\bibnamefont {Schuld}}, \bibinfo
  {author} {\bibfnamefont {C.}~\bibnamefont {Gogolin}}, \bibinfo {author}
  {\bibfnamefont {S.}~\bibnamefont {Ahmed}}, \bibinfo {author} {\bibfnamefont
  {V.}~\bibnamefont {Ajith}}, \bibinfo {author} {\bibfnamefont {M.~S.}\
  \bibnamefont {Alam}}, \bibinfo {author} {\bibfnamefont {G.}~\bibnamefont
  {Alonso-Linaje}}, \bibinfo {author} {\bibfnamefont {B.}~\bibnamefont
  {AkashNarayanan}}, \bibinfo {author} {\bibfnamefont {A.}~\bibnamefont
  {Asadi}}, \emph {et~al.},\ }\bibfield  {title} {\bibinfo {title} {Pennylane:
  Automatic differentiation of hybrid quantum-classical computations},\
  }\href@noop {} {\bibfield  {journal} {\bibinfo  {journal} {arXiv preprint
  arXiv:1811.04968}\ } (\bibinfo {year} {2018})}\BibitemShut {NoStop}%
\bibitem [{\citenamefont {Sim}\ \emph {et~al.}(2019)\citenamefont {Sim},
  \citenamefont {Johnson},\ and\ \citenamefont
  {Aspuru-Guzik}}]{sim2019expressibility}%
  \BibitemOpen
  \bibfield  {author} {\bibinfo {author} {\bibfnamefont {S.}~\bibnamefont
  {Sim}}, \bibinfo {author} {\bibfnamefont {P.~D.}\ \bibnamefont {Johnson}},\
  and\ \bibinfo {author} {\bibfnamefont {A.}~\bibnamefont {Aspuru-Guzik}},\
  }\bibfield  {title} {\bibinfo {title} {Expressibility and entangling
  capability of parameterized quantum circuits for hybrid quantum-classical
  algorithms},\ }\href@noop {} {\bibfield  {journal} {\bibinfo  {journal}
  {Advanced Quantum Technologies}\ }\textbf {\bibinfo {volume} {2}},\ \bibinfo
  {pages} {1900070} (\bibinfo {year} {2019})}\BibitemShut {NoStop}%
\bibitem [{\citenamefont {Brunton}\ \emph {et~al.}(2020)\citenamefont
  {Brunton}, \citenamefont {Noack},\ and\ \citenamefont
  {Koumoutsakos}}]{brunton2020machine}%
  \BibitemOpen
  \bibfield  {author} {\bibinfo {author} {\bibfnamefont {S.~L.}\ \bibnamefont
  {Brunton}}, \bibinfo {author} {\bibfnamefont {B.~R.}\ \bibnamefont {Noack}},\
  and\ \bibinfo {author} {\bibfnamefont {P.}~\bibnamefont {Koumoutsakos}},\
  }\bibfield  {title} {\bibinfo {title} {Machine learning for fluid
  mechanics},\ }\href@noop {} {\bibfield  {journal} {\bibinfo  {journal}
  {Annual review of fluid mechanics}\ }\textbf {\bibinfo {volume} {52}},\
  \bibinfo {pages} {477} (\bibinfo {year} {2020})}\BibitemShut {NoStop}%
\bibitem [{\citenamefont {Fukami}\ \emph {et~al.}(2020)\citenamefont {Fukami},
  \citenamefont {Fukagata},\ and\ \citenamefont
  {Taira}}]{fukami2020assessment}%
  \BibitemOpen
  \bibfield  {author} {\bibinfo {author} {\bibfnamefont {K.}~\bibnamefont
  {Fukami}}, \bibinfo {author} {\bibfnamefont {K.}~\bibnamefont {Fukagata}},\
  and\ \bibinfo {author} {\bibfnamefont {K.}~\bibnamefont {Taira}},\ }\bibfield
   {title} {\bibinfo {title} {Assessment of supervised machine learning methods
  for fluid flows},\ }\href@noop {} {\bibfield  {journal} {\bibinfo  {journal}
  {Theoretical and Computational Fluid Dynamics}\ }\textbf {\bibinfo {volume}
  {34}},\ \bibinfo {pages} {497} (\bibinfo {year} {2020})}\BibitemShut
  {NoStop}%
\end{thebibliography}%
\end{document}